\documentclass[11pt]{article}

\usepackage{amsthm}
\usepackage{bbm}
\usepackage{graphicx}
\usepackage{epstopdf}
\usepackage{amsfonts}
\usepackage[colorlinks,citecolor=blue,urlcolor=blue]{hyperref}
\usepackage{amsmath}
\usepackage{amstext}
\newtheorem{theorem}{Theorem}[section]
\newtheorem{lemma}[theorem]{Lemma}
\newtheorem{proposition}[theorem]{Proposition}
\newtheorem{corollary}[theorem]{Corollary}
\newtheorem{remark}[theorem]{Remark}
\newtheorem{definition}[theorem]{Definition}
\newtheorem{assumption}[theorem]{Assumption}
\newtheorem{example}[theorem]{Example}

\newcommand{\rulex}{\hfill\rule{1mm}{3mm}}
\begin{document}

\title{Continuous-time mean-variance portfolio selection under non-Markovian regime-switching model with random horizon}

\author{CHEN, Tian\thanks{Zhongtai Securities Institute for Financial Studies, Shandong University, Jinan $250100$, China. Email: 201812073@mail.sdu.edu.cn.}
\and
LIU, Ruyi\thanks{School of Mathematics and Statistics, University of Sydney, Sydney NSW $2006$, Australia. Email: ruyi.liu@sydney.edu.au.}
\and
WU, Zhen\thanks{School of Mathematics, Shandong University, Jinan $250100$, China.  Email: wuzhen@sdu.edu.cn.}}

\maketitle

\date{}

\begin{abstract} In this paper, we consider a continuous-time mean-variance portfolio selection with regime-switching and random horizon. Unlike previous works, the dynamic of assets are described by non-Markovian regime-switching models in the sense that all the market parameters are predictable with respect to the filtration generated jointly by Markov chain and Brownian motion. We formulate this problem as a constrained stochastic linear-quadratic optimal control problem. The Markov chain is assumed to be independent of the Brownian motion. So the market is incomplete. We derive closed-form expressions for both the optimal portfolios and the efficient frontier. All the results are different from those in the problem with fixed time horizon.      

\bigskip\noindent
{\bf Key words:} Mean-variance portfolio selection, random time horizon, stochastic LQ control, backward stochastic differential equation        

\bigskip\noindent
{\bf MR(2000) Subject Classification:} 05B05, 05B25, 20B25

\end{abstract}
\section{Introduction}

Mean-variance portfolio selection problem is concerned about the tradeoff between the terminal return and the associated risk of the investment among a number of securities. It was first proposed and solved in the single-period setting by Markowitz \cite{Markowitz1952}. Zhou and Li \cite{Zhou2000} studied a continuous-time mean-variance problem and introduced an appropriate, effective framework in terms of stochastic LQ controls for this problem. In their paper, all the market parameters were assumed to be deterministic. Motivated by the need of more practical models, Lim and Zhou \cite{Lim2002} solved a mean-variance problem with random market parameters in a complete market. Zhou and Yin \cite{Zhou2003} studied the continuous-time mean-variance portfolio selection with regime-switching, they illustrated the influence of market trend change on portfolio selection by virtue of Markov chain. Lim \cite{Lim2004, Lim2005} studied the same problem in an incomplete market or with jumps, respectively. Based on the work of Lim \cite{Lim2004} and Yu \cite{Yu2013},  Lv et al. \cite{Lv2016} studied the mean-variance problem with random horizon in an incomplete market. Meanwhile, they proved that BMO martingales can be used to deal with the stochastic Riccati equations and the auxiliary BSDEs arising from various mean-variance problems. Shen et al. \cite{Shen2020} studied a mean-variance asset-liability management problem under non-Markovian regime-switching models. In their paper, Markov chain is no longer explicitly contained in the market parameters, but the model parameters are predictable with respect to the filtration generated by Markov chain and Brownian motion.

Most of classical financial economics are based on the hypothesis that the investors know with certainty about the eventual time of exit when they make investment decisions. But in practice, when many investors enter the market, they are not sure when they will exit. Therefore, it is of great significance to develop financial theory in random time horizon. Yaari \cite{Yaari1965} studied an optimal consumption problem for an individual with uncertain lifetime at the beginning. Then Hakansson \cite{Hakansson1969,Hakansson1971}, Richard \cite{Richard1975} etc. assumed that the random exit time is independent of all other uncertainties in the market. On the contrary, under the condition that the random exit time is fully dependent on the prices of underlying assets, Karatzas and Wang \cite{Karatzas2000} solved an optimal dynamic investment problem in a complete market. Bouchard and Pham \cite{Bouchard2004} and Blanchet-Scalliet et al. \cite{Blanchet2008} extended these two extreme cases to the situation that the horizon terminal is a general random time. By assuming all market parameters and the conditional distribution function of exit time were deterministic in \cite{Blanchet2008}, the authors obtained the explicit form of optimal portfolio for the constant relative risk aversion (CRRA) utility, which coincides with the outcomes under fixed time horizon. Then Yu \cite{Yu2013} and Lv et al. \cite{Lv2016} studied the mean-variance portfolio selection with random horizon in the complete market and incomplete market respectively. Huang, Wang and Wu \cite{Huang2020} studied a kind of optimal investment problem under inflation and uncertain time horizon. Wang and Wu \cite{Wang2020} studied mean-variance portfolio selection with discontinuous prices and random horizon in an incomplete market.

In this paper, we study a continuous-time mean-variance portfolio selection problem under non-Markovian regime-switching model with a general random time horizon and derive the closed form expressions for efficient portfolios and efficient frontier. In order to be more practical, we assume that all market parameters are predictable and the exit time is random. It is not a stopping time, which means that the exit time depends not only on the price information, but also on other uncertain factors in the market. This random horizon in deterministic case is first introduced by Blanchet-Scalliet et al. \cite{Blanchet2008} and we extend it to the stochastic case with Markov chain. We use a submartingale to characterize the conditional distribution of random time $\tau$ and reconstruct the mean-variance problem according to the Doob-Mayer decomposition theorem and some assumptions.

When we apply the LQ approach to the mean-variance problem, the key difficulty is to prove the global solvability of the so-called stochastic Riccati equation and the auxiliary regime-switching BSDE arising from the problem. When the time horizon is deterministic and the market parameters are random, however the SRE is a fully nonlinear singular BSDE for which the usual assumption (such as the Lipschitz and linear growth conditions) are not satisfied. Fortunately, the SRE has a nice structure. By BMO martingale and Girsanov theorem, Shen et al. \cite{Shen2020} obtained the existence and uniqueness of the solution of the SRE in the case. When the time horizon and the market parameters are both random which is the case concerned in this paper, the corresponding SRE is more complicated. In detail, there is an additional item  destroying the nice structure. In this paper, we use a truncation technique to transform the SRE into a one-dimensional BSDE with quadratic grwoth. With the help of BMO martingale technique and comparison theorem and the result of Shen et al. \cite{Shen2020}, we obtain the existence and uniqueness of the stochastic Riccati equation and the auxiliary regime-switching BSDE. Then we get the efficient portfolios in a feedback form as well as the efficient frontier. In fact, the solutions of the two BSDEs completely determine the efficient portfolios and efficient frontier of the underlying mean-variance problem. In addition, we also derive the minimum variance explicitly. Because of the influence of uncertain exit time and Markov chain, the efficient frontier is no longer a perfect square (different from Lim and Zhou \cite{Lim2002}). As a result, the investors are not able to achieve a risk-free investment.

The rest of this paper is organized as follows. In section 2, we construct the non-Markovian regime-switching model with a random time horizon and formulate the corresponding mean-variance problem. In section 3, we investigated the feasibility property of the underlying model. In section 4, we show the global solvability of the stochastic Riccati equation and the auxiliary regime-switching BSDE. In section 5, we derive the solution of the unconstrained optimization problem. In section 6, we present the efficient portfolios and efficient frontier. Finally, we present the conclusion.

\section{Problem Formulation}

Let $\left({\it \Omega}, \mathcal{A}, \mathbb{F}, \mathbf{P} \right)$ be a complete filtered probability space. $W(t) = \left(W_1(t), W_2(t), \cdots, W_n(t) \right)^{\rm T}$ is an $\mathbb{R}^n$ valued standard Brownian motion (with $W(0) = \mathbf{0}$), where the superscript $M^{\rm T}$ denotes the transpose of any vector or matric $M$. And $\alpha(t)$ denotes a continuous-time finite state homogeneous Markov chain on this probability space.
We assume that the Brownian motion $W(\cdot)$ and the Markov chain $\alpha(\cdot)$ are independent. We further assume that $T > 0$ is a fixed time horizon and the filtration $\mathbb{F} = \{\mathcal{F}_{t}: 0 \leq t \leq T \}$ with ${\mathcal{F}}_T \subset \mathcal{A}$ is generated by $W(\cdot)$ and $\alpha(\cdot)$,
\begin{equation*}
  \mathcal{F}_t := \sigma\{W(s), \alpha(s): 0 \leq s \leq t\} \vee \mathcal{N}({\mathbf{P}}),\quad\quad \forall t\in[0,T],
\end{equation*}
where $\mathcal{N}(\mathbf{P})$ denotes the collection of all $\mathbf{P}$-null events in this probability space so that $t \mapsto \mathcal{F}_t$ is continuous.

Throughout this paper, the state space of the Markov chain is identified with the canonical state space, that is, a finite set of unit vectors $\mathcal{E}:=\{e_1,e_2,\cdots,e_N\}\subset\mathbb{R}^N$, where the $j$th component of $e_i$ is the Kronecker delta $\delta_{ij}$, for each $i,j=1,2,\cdots, N$. We assume that the Markov chain $\alpha$ has a generator $Q=(q_{ij})_{N\times N}$ and stationary transition probabilities
\begin{equation}\label{transitionp}
  p_{ij}(t)=\mathbf{P}(\alpha(t)=e_j | \alpha(0)=e_i), \quad t\geq 0,\ i,j=1,2,\dots,N.
\end{equation}
Since $q_{ij}\geq 0$, for $i\ne j$ and $\sum_{i=1}^N q_{ij}=0$, we have $q_{ii}<0$, for each $i=1,2,\cdots, N$. Based on the canonical representation of the state space, Elliott et al. \cite{Elliott1995} provided the following semimartingale representation of the chain $\alpha$:
\begin{equation*}
  \alpha(t)=\alpha(0)+\int_{0}^{t} Q'\alpha(u)\mathrm{d}u +\textbf{M}(t),
\end{equation*}
where $\{\textbf{M}(t)| t\in [0,T] \}$ is an $\mathbb{R}^N$-valued, $(\mathbb{F}, \mathbf{P})$-martingale.

For each $i,j=1,2,\cdots,N$, with $i\ne j$, and $t\in [0,T]$, let $J^{ij}(t)$ be the number of jumps from state $e_i$ to state $e_j$ up to time $t$. Then
\begin{equation*}
  \begin{aligned}
    J^{ij}(t):=&\sum_{0<s\leq t}\langle\alpha(s-),e_i\rangle\langle\alpha(s),e_j\rangle=\sum_{0<s\leq t}\langle\alpha(s-),e_i\rangle \langle\alpha(s)- \alpha(s-),e_j\rangle\\
    =&\int_0^t \langle\alpha(s-),e_i\rangle\langle\mathrm{d}\alpha(s),e_j\rangle=q_{ij}\int_0^t \langle\alpha(s-),e_i\rangle\mathrm{d}s+m_{ij}(t),
  \end{aligned}
\end{equation*}
where $m_{ij}(t):=\int_0^t\langle\alpha(s-),e_i\rangle\langle \mathrm{d} \mathbf{M}(s), e_j\rangle$ is $(\mathbb{F},\mathbf{P})$-martingale. $\langle\cdot, \cdot \rangle$ denotes the inner product of the Euclidean space. The $m_{ij}$ are called the basic martingales associated with the chain $\alpha$.

Now, for each fixed $j=1,2,\cdots, N$, let $\Phi_j(t)$ be the number of jumps into state $e_j$ up to time $t$. Then
\begin{equation*}
  {\it \Phi}_j(t):=\sum_{i=1,i\ne j}^N J_{ij}(t)=\sum_{i=1,i\ne j}^N q_{ij}\int_0^t\langle \alpha(s),e_i \rangle \mathrm{d}s +{\it \tilde{\Phi}}_j(t),
\end{equation*}
where $\tilde{\it \Phi}_j(t):=\sum_{i=1,i\ne j}^N m_{ij}(t)$ is again an $(\mathbb{F},\mathbf{P})$-martingale  for each $j=1,2,\cdots, N$.

For each $j=1,2,\cdots, N$, let denote the compensator of ${\it \Phi}(t)$
\begin{equation*}
  \lambda_j(t):=\sum_{i=1,i\ne j}^N q_{ij}\langle \alpha(s),e_i\rangle,
\end{equation*}
such that
\begin{equation*}
  \tilde{\it \Phi}_j(t)={\it \Phi}_j(t)-\int_0^t \lambda_j(s) \mathrm{d}s.
\end{equation*}
To simplify our notation, we denote the vector of counting processes $\left\{{\it \Phi}(t)|t\in [0,T]\right\}$, intensity processes $\left\{\lambda(t)|t\in[0,T]\right\}$ and compensated counting processes $\{\tilde{\it \Phi}(t)|t\in [0,T]\}$ by ${\it \Phi}(t):=({\it \Phi}_1(t), {\it \Phi}_2(t), \cdots, {\it \Phi}_N(t))', \lambda(t):=(\lambda_1(t),\lambda_2(t),\dots, \lambda_N(t))'$ and $\tilde{\it \Phi}(t):= (\tilde{\it \Phi}_1(t), \tilde{\it \Phi}_2(t), \cdots,\tilde{\it \Phi}_N(t))'$, respectively, and these three processes are related as
$$\tilde{\it \Phi}(t)={\it \Phi}(t)-\int_0^t\lambda(s)\mathrm{d}s, \quad t\in [0,T].$$

Furthermore, it is worthwhile to point out that the jump of ${\it \Phi}(t)$, at any $t\in [0,T]$, can only take values on the set $\{0_N,e_1,e_2,\dots, e_N\}$, where $0_N$ is an $N$ dimensional vector of zeros. To be more precise, ${\it \Delta  \Phi}(t)=e_j$, if there is a transition into state $e_j$ from any one of the other states at time $t$; otherwise, ${\it \Delta\Phi}(t)=0_N$, if there is no transition at all states at time $t$.

Now we introduce some space of stochastic processes and random variables.
\begin{itemize}

\item $L^2_{\mathbb{F}}(0,T;\mathbb{R}^m)$, the set of $\mathbb{R}^m$-valued $\mathbb{F}$-predictable processes $\varphi$ defined on $[0,T]$ such that
    $$\mathbb{E}\int_0^T |\varphi(t)|^2 \mathrm{d}t < \infty;$$
\item ${\it\Pi}_{\mathbb{F}}^2(0,T;\mathbb{R}^N)$, the set of $\mathbb{R}^N$-valued $\mathbb{F}$-predictable processes $\varphi$ defined on $[0,T]$ such that
    $$\mathbb{E}\int_0^T |\varphi(t)|^2\mathrm{d}{\it\Phi}(t)<\infty;$$
\item $L^{2,loc}_{\mathbb{F}}(0,T;\mathbb{R}^m)$, the set of $\mathbb{R}^m$-valued $\mathbb{F}$-predictable processes $\varphi$ defined on $[0,T]$ such that
    $$\int_0^T |\varphi(t)|^2\mathrm{d}t < \infty \qquad \mathbf{P}-a.s.;$$
\item $L^{\infty}_{\mathbb{F}}(0,T;\mathbb{R}^m)$, the set of $\mathbb{F}$-adapted, uniformly bounded processes;
\item $S^2_{\mathbb{F}}(0,T;\mathbb{R}^m)$, the set of $\mathbb{R}^m$-valued $\mathbb{F}$-adapted, continuous processes $\varphi$ defined on $[0,T]$ such that
    $$\mathbb{E}\left[\sup_{t\in[0,T]}|\varphi(t)|^2\right] < \infty ;$$
\item $S^{\infty}_{\mathbb{F}}(0,T;\mathbb{R}^m)$, the set of $\mathbb{F}$-adapted, uniformly bounded, continuous processes;
\item $L^{\infty}(\Omega, \mathcal{F}_T, \mathbf{P}; \mathbb{R}^m)$, the set of $\mathbb{R}^m$-valued, $\mathcal{F}_T$-measurable, bounded random variables;
\item ${\rm BMO}_{\mathbf{P}}(0,T;\mathbb{R})$, the space of $\mathbb{R}$-valued, {\rm BMO} martingales on $(\mathbb{F},\mathbf{P})$ equipped with the norm
    $$\|\varphi\|_{{\rm BMO_{\mathbf{P}}(0,T;\mathbb{R})}}:=\sup_{\tau\in\mathcal{T}}\left\|\left\{\mathbb{E}\left[ |\varphi(T)-\varphi(\tau{-})|^2|\mathcal{F}_{\tau}\right]\right\}^{\frac{1}{2}}\right\|_{\infty}<\infty,$$
    where $\mathcal{T}$ is the set of all $\mathbb{F}$ stopping time on $[0,T]$. Whenever there is no confusion, we abbreviate ${\rm BMO}_{\mathbf{P}}(0,T;\mathbb{R})$ as ${\rm BMO}_{\mathbf{P}}$. More details of BMO martingales can be seen in \cite{Kazamaki2006}.
\item $\mathcal{H}^2_{{\rm BMO}_{\mathbf{P}}}(0,T;\mathbb{R}^n)$, the space of $\mathbb{R}^n$-valued, $\mathbb{F}$-predictable processes $\varphi$ such that $\int_0^{\cdot}\varphi(t)^{\rm T}\mathrm{d}W(t)\\
    \in {\rm BMO}_{\mathbf{P}}$, equipped with the norm
    $$\|\varphi\|_{\mathcal{H}^2_{{\rm BMO}_{\mathbf{P}}}(0,T;\mathbb{R}^n)}:=\left\|\int_0^{\cdot}\varphi(t)^{\rm T}   \mathrm{d}W(t) \right\|_{{\rm BMO}_{\mathbf{P}}}=\sup_{\tau\in\mathcal{T}}\left\|\left\{\mathbb{E}\left[\int_{\tau}^T|\varphi(t)|^2 \mathrm{d}t|\mathcal{F}_{\tau}\right]\right\}^{\frac{1}{2}}\right\|_{\infty}<\infty.$$
\item $\mathcal{J}^2_{{\rm BMO}_{\mathbf{P}}}(0,T;\mathbb{R}^N)$, the space of $\mathbb{R}^N$-valued, $\mathbb{F}$-predictable processes $\varphi$ such that $\int_0^{\cdot}\varphi(t)^{\rm T}\mathrm{d}\tilde{\Phi} (t)\\
    \in {\rm BMO}_{\mathbf{P}}$, equipped with the norm
    \begin{equation*}
      \begin{aligned}
        \|\varphi\|_{\mathcal{J}^2_{{\rm BMO}_{\mathbf{P}}}(0,T;\mathbb{R}^N)}:&=\left\|\int_0^{\cdot}\varphi(t)^{\rm T} \mathrm{d}\tilde{\it\Phi}(t) \right\|_{{\rm BMO}_{\mathbf{P}}}\\
        &=\sup_{\tau\in\mathcal{T}}\left\|\left\{\mathbb{E}\left[\sum_{j=1}^N\int_{\tau}^T |\varphi(t)|^2 \mathrm{d}{\it\Phi}_j(t)\bigg|\mathcal{F}_{\tau}\right]\right\}^{\frac{1}{2}}\right\|_{\infty}<\infty.
      \end{aligned}
    \end{equation*}
\end{itemize}

Consider a market with $n+1$ securities, consisting of a bond and $n$ stocks are traded continuously. The bond price $P_0(t)$ satisfies the following equation:
\begin{equation}\label{bondprice}
  \begin{cases}
    \mathrm{d}P_0(t) = r\left(t\right)P_0(t) \mathrm{d}t, \quad t \in [0,T],\\
    P_0(0) = p_0 > 0,
  \end{cases}
\end{equation}
where $r(t)$ is given as the interest rate process. The price of each stock $P_1(t), P_2(t), \cdots, P_n(t)$, satisfies the stochastic differential equation:
\begin{equation}\label{stockprice}
\left\{
  \begin{aligned}
    &\mathrm{d}P_m(t) = P_m(t)\left\{\mu_m\left(t\right)\mathrm{d}t + \sum_{j=1}^n \sigma_{mj}\left(t\right) \mathrm{d}W_j(t)\right\}, \qquad t \in [0,T],\\
    &P_m(0)=p_m > 0,
  \end{aligned}
  \right.
\end{equation}
where $\mu_m(t)$ is the appreciation rate process and $\sigma_m(t) := \big(\sigma_{m1}(t), \sigma_{m2}(t), \cdots,  \sigma_{mn}(t)\big)$ is the volatility (or dispersion) rate process of the $m$th stock. Then define the volatility matrix $\sigma(t):=(\sigma_{mj}(t))_{n\times n}$.


\begin{assumption}\label{2.1}
  The interest rate process $\{r(t)|t\in[0,T]\}$ is a positive, uniformly bounded, $\mathbb{F}$-predictable process; the appreciation rate processes $\{\mu_m(t)|t\in [0,T]\}$ for $m=1,2,\cdots,n$, are uniformly bounded, $\mathbb{F}$-predictable processes; the volatility processes $\{\sigma_{mj}(t)|t\in [0,T]\}$, for $m=1,2,\cdots,n$ and $j=1,2,\cdots,n$ are nonnegative, uniformly bounded, $\mathbb{F}$-predictable processes, and satisfy the non-degenerate condition, that is, for any $t\in [0,T]$, there exists a positive constant $\delta$ such that $\sigma(t)\sigma(t)^{\rm T} \geq \delta I_{n\times n}$.
\end{assumption}

\begin{remark}
  In our model, all market parameters are assumed to be predictable with respect to the filtration generated jointly by Markov chain and Brownian motion. That is, they are completely stochastic rather than explicitly containing Markov chains. So our model is called non-Markovian regime-switching model.

\end{remark}


 Consider an agent who invests the amount $\pi_m(t)$ of the wealth $x(t)$ in the $m$th stock ($m=1,2,\dots,n$). If the strategy $\pi(t)=\left( \pi_1(t), \pi_2(t), \cdots, \pi_n(t)\right)^{\rm T}$ is self-financing, the wealth invested in the riskless asset is $x(t) - \sum^n_{m=1} \pi_m(t)$, then the wealth process $x(\cdot)$, with the initial endowment $x_0$, satisfies the following SDE
\begin{equation}\label{wealths}
  \left\{
  \begin{aligned}
    &\mathrm{d}x(t) =\left\{r(t)x(t)+\sum_{m=1}^n \left[\mu_m(t)-r(t)\right]\pi_m(t) \right\}\mathrm{d}t
    + \sum_{m=1}^n\pi_m(t) \sum_{j=1}^n \sigma_{mj}(t)\mathrm{d}W_j(t),\\
    &x(0)=x_0>0.
  \end{aligned}
  \right.
\end{equation}
Setting
\begin{equation}
  B(t) := \left(\mu_1(t)-r(t),\mu_2(t)-r(t),\cdots,\mu_n(t)-r(t)\right)^{\rm T},
\end{equation}
we can rewrite the wealth equation \eqref{wealths} as

\begin{equation}\label{wealth}
  \left\{
  \begin{aligned}
    &\mathrm{d}x(t)=\left[r(t)x(t)+B(t)^{\rm T}\pi(t)\right]\mathrm{d}t+\pi(t)^{\rm T}\sigma(t) \mathrm{d}W(t)\\
    &x(0)=x_0>0.
  \end{aligned}
  \right.
\end{equation}

\begin{definition}\label{Def2.3}
  $\pi(t)$ is called an admissible portfolio if and only if $\pi(\cdot)\in \mathcal{U}:=L^2_{\mathbb{F}}(0,T;\mathbb{R}^n)$. It is easy to see that, for any admissible $\pi(\cdot)$, the corresponding SDE \eqref{wealth} has a unique solution $x(\cdot)\in S^2_{\mathbb{F}}(0,T;\mathbb{R})$ . In this case, we refer to $(x(\cdot),\pi(\cdot))$ as an admissible pair.
\end{definition}

\begin{remark}
  In many literature, a portfolio named $u(\cdot)$ is defined as the proportion of wealth allocated to different stocks, i.e.
  \begin{equation}\label{positive control}
    u(t)=\frac{\pi(t)}{x(t)},\qquad t\in [0,T].
  \end{equation}
\end{remark}

By this definition, the wealth equation \eqref{wealth} can be written as
\begin{equation}
  \left\{
  \begin{aligned}
    &\mathrm{d}x(t)=x(t)\left\{\left[r(t)+B(t)^{\rm T}u(t)\right]\mathrm{d}t+u(t)^{\rm T}\sigma(t)
    \mathrm{d}W(t)\right\},\\
    &x(0)=x_0.
  \end{aligned}
  \right.
\end{equation}
It is easy to know that the above equation admits a unique positive solution when the initial endowment $x_0$ is positive from the standard SDE theory. In other words, with the definition of portfolio \eqref{positive control}, the corresponding wealth process must be automatic positive. However, the wealth process with zero or negative values is also sensible at least for some circumstance (see Zhou and Yin \cite{Zhou2003}).


In reality, the investor cannot know when the investment will exit certainly. We assume that the exit time of the investment is $\tau \wedge T$, where $\tau$ is a $\mathcal{A}$-measurable positive random variable. Under our assumption, $\mathcal{A}$ may be strictly bigger than $\mathcal{F}_T$. It means that the exit time relies on not only the uncertainties of prices, but also other uncertain factors in the market.

\begin{remark}
  It is easy to see that $\tau$ is not a stopping time under our assumption.
\end{remark}

In this paper, we use the formulation of random time horizon similar to Blanchet-Scalliet et al. \cite{Blanchet2008}. We introduce the conditional distribution function of exit time by $F(t)=\mathbf{P} (\tau \leq t|\mathcal{F}_t)$. It is easy to verify that $F(t)$ is an $\mathbb{F}$-submartingale and the function $t\rightarrow \mathbb{E}[F(t)]$ is right-continuous. Then $F(t)$ has a right-continuous modification. From the Doob-Mayer decomposition theorem, we have $F(t)=A(t)+M(t)$, where $M$ is a martingale and $A$ is an increasing process. Then we give the following assumptions.

\begin{assumption}\label{2.7}
  The process $A(\cdot)$ is absolutely continuous with respect to Lebesgue's measure, with a nonnegative bounded density $f(\cdot)$, i.e., $A(t)=\int_0^t f(s) \mathrm{d}s,\ t\in [0,T]$.
\end{assumption}

Under Assumption \ref{2.7}, we immediately get the boundedness of $A$, and then of the martingale $M$. From the martingale representation theorem, there exist processes $Z(\cdot) \in L^2_{\mathbb{F}}(0,T;\mathbb{R}^n)$ and $K(\cdot)\in {\it\Pi}_{\mathbb{F}}^2(0,T;\mathbb{R}^N)$ such that
$$M(t)=\int_0^t Z(s) \mathrm{d}W(s)+\int_0^t K(s) \mathrm{d}\tilde{\it \Phi}(s),\qquad t\in [0,T].$$
Actually, by the boundedness of $M$ and the Burkholder-Davis-Gundy inequality, there exists a constant $c_2$ such that
\begin{equation*}
  \begin{aligned}
    &\mathbb{E}\left[\int_0^T |Z(t)|^2 \mathrm{d}t\right]+\mathbb{E}\left[\int_0^T |K(t)|^2 \mathrm{d}{\it\Phi}(t)\right]\leq c_2\mathbb{E}\left[\sup_{t\in[0,T]}|M(t)|^2\right]<\infty.\\
  \end{aligned}
\end{equation*}
But, we strengthen the above conclusion to the following

\begin{assumption}\label{2.8}
  There exists a constant $C$ such that
  \begin{equation*}
    \int_0^T|Z(t)|^2\mathrm{d}t + \int_0^T|K(t)|^2\mathrm{d}{\it\Phi}(t) \leq C,\qquad \mathbf{P}-a.s..
  \end{equation*}
\end{assumption}

\begin{assumption}\label{2.9}
  There exists a constant $\varepsilon > 0$ such that $F(T)\leq 1-\varepsilon,\ \mathbf{P}-a.s.$.
\end{assumption}

\begin{remark}
  Blanchet-Scalliet et al. \cite{Blanchet2008} assumed that the martingale $M\equiv 0$. Here, we weaken this condition.
\end{remark}

The agent's objective is to find an admissible portfolio $\pi(\cdot)$, among all such admissible portfolios whose expected terminal wealth $\mathbb{E}[x(\tau \wedge T)]=z$, for some given $z\in \mathbb{R}$, so that the risk measured by the variance of the terminal wealth
\begin{equation}
  {\rm Var } [x(\tau \wedge T)]:=\mathbb{E}\big\{x(\tau \wedge T)-\mathbb{E}[x(\tau \wedge T)]\big\}^2 = \mathbb{E}[x(\tau\wedge T)-z]^2
\end{equation}
is minimized. Finding such a portfolio $\pi(\cdot)$ is referred to as the mean-variance portfolio selection problem with a random horizon. 

From the definition of $F(t)=\mathbf{P}(\tau \leq t|\mathcal{F}_t)$, its decomposition $F(t)=A(t)+M(t)$ and Assumptions \ref{2.7}-\ref{2.9}, we have the following formulation by the same argument of Lemma 2.5 in Yu \cite{Yu2013},
\begin{equation*}
  \begin{aligned}
    \mathbb{E}[x(\tau \wedge T)] &=\mathbb{E}[\mathbbm{1}_{\{\tau\leq T\}}x(\tau)+\mathbbm{1}_{\{\tau >T\}}x(T)]
    =\mathbb{E}\left[\int_0^T x(t) \mathrm{d}F(t)+\int_T^{\infty} x(T)\mathrm{d}F(t)\right]\\
    &=\mathbb{E}\left[\int_0^T f(t)x(t)\mathrm{d}t+(1-F(T))x(T)\right].
  \end{aligned}
\end{equation*}

Similarly, noting that $z=\mathbb{E}[x(\tau \wedge T)]$, we have
\begin{equation*}
  {\rm Var } [x(\tau \wedge T)] = \mathbb{E} \left[\int_0^T f(t)(x(t)-z)^2\mathrm{d}t + (1-F(T))(x(T)-z)^2\right].
\end{equation*}

\begin{definition}\label{Del2.11}
  Under Assumptions \ref{2.7}-\ref{2.9}, the mean-variance portfolio selection problem with random time horizon is formulated as a linearly constrained stochastic optimization problem, parameterized by $z \in \mathbb{R}$:
  \begin{equation}\label{question}
    \left\{
    \begin{aligned}
      &{\rm minimize}\  J_{MV}(\pi(\cdot)):=\mathbb{E}\left[\int_0^T f(t)(x(t)-z)^2\mathrm{d}t + (1-F(T))(x(T)-z)^2\right],\\
      &{\rm subject\  to}\
      \left\{
      \begin{aligned}
      &J_1(\pi(\cdot)):=\mathbb{E}\left[\int_0^T f(t)x(t)\mathrm{d}t+(1-F(T))x(T)\right]=z,\\
      &(x(\cdot),\pi(\cdot)) \quad {\rm admissible}.
      \end{aligned}
      \right.
    \end{aligned}
    \right.
  \end{equation}
\end{definition}

Moreover, an admissible portfolio $\pi(\cdot)\in \mathcal{U}$ is said to be a feasible portfolio for problem \eqref{question} if it satisfies the constraint $J_1(\pi(\cdot))=z$. If there exists a feasible portfolio, then problem \eqref{question} is said to be feasible. Problem \eqref{question} is called finite if it is feasible and the infimum of $J_{MV}(\pi(\cdot))$ over the set of feasible portfolios is finite. If problem \eqref{question} is finite and the infimum of $J_{MV}(\pi(\cdot))$ is achieved by a feasible portfolio $\pi^*(\cdot)$, then problem \eqref{question} is said to be solvable and $\pi^*(\cdot)$ is called an optimal portfolio. Finally, an optimal portfolio to problem \eqref{question} is also called an efficient portfolio corresponding to $z$, and the corresponding pairs $({\rm Var } [x(\tau \wedge T)],z)\in \mathbb{R}^2$ and $(\sigma_{x(\tau\wedge T)},z)\in \mathbb{R}^2$ are interchangeably called an efficient point, where $\sigma_{x(\tau\wedge T)}=\sqrt{{\rm Var } [x(\tau \wedge T)]}$ denotes the standard deviation of $x(\tau \wedge T)$. The set of all efficient points is called the efficient frontier.

\section{Feasibility}

Since the problem \eqref{question} involves a linear constraint $J_1(\pi(\cdot))=z$, in this section, we derive the conditions under which the problem is at least feasible. In fact, we have the following result on the feasibility of problem \eqref{question}.

\begin{proposition}\label{Prop3.1}
  Let $\big(\psi(\cdot),\xi(\cdot),\hat{\xi}(\cdot)\big)\in S_{\mathbb{F}}^2(0,T;\mathbb{R})\times L_{\mathbb{F}}^2(0,T;\mathbb{R}^n)\times {\it\Pi}_{\mathbb{F}}^2(0,T;\mathbb{R}^N)$ be the solution of the following regime-switching BSDE:
  \begin{equation}\label{feasible}
  \left\{
    \begin{aligned}
      &-\mathrm{d}\psi(t)=\left[\psi(t)r(t)+f(t)\right]\mathrm{d}t-\xi(t)^{\rm T}\mathrm{d}W(t)- \hat{\xi}(t)^{\rm T}\mathrm{d}\tilde{\it\Phi}(t)\\
      &\psi(T)=1-F(T).
    \end{aligned}
    \right.
  \end{equation}
  Then the mean-variance problem \eqref{question} is feasible for every $z\in\mathbb{R}$ if and only if
  \begin{equation}\label{ftiao}
    \gamma:=E\int_0^T |\psi(t)B(t)+\xi(t)^{\rm T}\sigma(t)|^2 \mathrm{d}t > 0.
  \end{equation}
\end{proposition}
\proof
  First, it is easy to see that the linear BSDE \eqref{feasible} admits a unique solution

  In order to prove the sufficiency, we construct a family of admissible portfolios $\pi^{(\beta)}(\cdot)=\beta\pi(\cdot)$ for all $\beta \in \mathbb{R}$, where
  \begin{equation}\label{pi}
    \pi(t)= \psi(t)B(t)+\xi(t)^{\rm T}\sigma(t).
  \end{equation}
  Let $x^{(\beta)}(\cdot)$ be the wealth process corresponding to $\pi^{(\beta)}(\cdot)$ for any $\beta$. By linearity of the wealth equation \eqref{wealth}, it follows that $x^{(\beta)}(t)=x^{(0)}(t)+\beta y(t)$, where $x^{(0)}(\cdot)$ satisfies
  \begin{equation}
    \begin{cases}
      \mathrm{d}x^{(0)}(t)=r(t)x^{(0)}(t)\mathrm{d}t, \quad t\in[0,T],\\
      x^{(0)}(0)=x_0,
    \end{cases}
  \end{equation}
  and $y(\cdot)$ satisfies
  \begin{equation}\label{yy}
    \begin{cases}
      \mathrm{d}y(t)=\left[r(t)y(t)+B(t)^{\rm T}\pi(t)\right]\mathrm{d}t+\pi(t)^{\rm T}\sigma(t) \mathrm{d}W(t), \quad t\in [0,T],\\
      y(0)=0,
    \end{cases}
  \end{equation}
  respectively. The problem \eqref{question} is feasible for any $z\in \mathbb{R}$ if there exists a $\beta \in \mathbb{R}$ such that
  \begin{equation*}
    \begin{aligned}
      z=J_1(\pi^{(\beta)}(\cdot)) &= \mathbb{E}\left[ \int_0^T f(t)x^{(0)}(t)\mathrm{d}t + (1-F(T))x^{(0)}(T) \right]\\
      &\quad\quad+\beta \mathbb{E}\left[\int_0^T f(t)y(t)\mathrm{d}t+(1-F(T))y(T) \right].
    \end{aligned}
  \end{equation*}
  In other words, problem \eqref{question} is feasible for any $z\in \mathbb{R}$ if
  \begin{equation*}
    \mathbb{E}\left[\int_0^Tf(t)y(t)\mathrm{d}t + (1-F(T))y(T) \right]\neq 0.
  \end{equation*}
  However, applying It\^o's formula to $\psi(t)y(t)$ on the interval $[0,T]$. Integrating from $0$ to $T$ and taking expectation, we have
  \begin{equation}\label{psiy}
    \begin{aligned}
      \mathbb{E}\left[\int_0^T f(t)y(t)\mathrm{d}t+(1-F(T))y(T)\right]
      &=\mathbb{E}\left\{\int_0^T \Big[\psi(t)B(t)+\xi(t)^{\rm T}\sigma(t)\Big]^{\rm T} \pi(t) \mathrm{d}t \right\}\\
      &=\mathbb{E}\left\{\int_0^T \Big[\psi(t)B(t)+\xi(t)^{\rm T}\sigma(t) \Big]^2  \mathrm{d}t \right\}.
    \end{aligned}
  \end{equation}
  Consequently, $\mathbb{E}y(T)\neq 0$ if \eqref{ftiao} holds.

  Conversely, if problem \eqref{question} is feasible for any $z\in \mathbb{R}$, then there exists an admissible portfolio $\pi(\cdot)$ such that
  \begin{equation*}
    \mathbb{E}\left[\int_0^T f(t)x(t)\mathrm{d}t + (1-F(T))x(T)\right]=z.
  \end{equation*}
  We can decompose $x(t)=x^{(0)}(t)+y(t)$, where $y(\cdot)$ satisfies \eqref{yy}. This leads to
  \begin{equation*}
    \mathbb{E}\left[\int_0^Tf(t)x^{(0)}(t)\mathrm{d}t+(1-F(T))x^{(0)}(T)\right] + \mathbb{E}\left[\int_0^Tf(t)y(t)\mathrm{d}t+(1-F(T))y(T)\right]=z.
  \end{equation*}
  However $x^{(0)}(\cdot)$, which can be interpreted as the wealth process corresponding to the agent putting all the money in the bond, is independent of $\pi(\cdot)$, thus it is necessary that there exists a $\pi(\cdot)$ such that
  \begin{equation*}
    \mathbb{E}\left[\int_0^T f(t)y(t)\mathrm{d}t + (1-F(T))y(T)\right]\neq 0.
  \end{equation*}
  From \eqref{psiy},
  \begin{equation*}
    \mathbb{E}\left\{\int_0^T \Big[\psi(t)B(t)+\xi(t)^{\rm T}\sigma(t) \Big]^{\rm T}\pi(t)\mathrm{d}t\right\}\neq 0.
  \end{equation*}
  This implies \eqref{ftiao}. \rulex

\begin{corollary}\label{Cor3.2}
  Let
  \begin{equation*}
    z^{(0)}=\mathbb{E}\left[\int_0^T f(t)x^{(0)}(t)\mathrm{d}t + (1-F(T))x^{(0)}(T)\right],
  \end{equation*}

  \begin{equation*}
    \gamma=E\int_0^T |\psi(t)B(t)+\xi(t)^{\rm T}\sigma(t)|^2 \mathrm{d}t.
  \end{equation*}
  We have:\\
  {\rm (i)} if \eqref{ftiao} holds, then for any $z\in \mathbb{R}$, a feasible portfolio satisfying $J_1(\tilde{\pi}(\cdot))=z$ is given by
  \begin{equation*}
    \tilde{\pi}(\cdot)=\frac{z-z^{(0)}}{\gamma}\left(\psi(t)B(t)+\xi(t)^{\rm T}\sigma(t)\right);
  \end{equation*}
  {\rm (ii)} if \eqref{ftiao} doesn't hold, then for any admissible portfolio $\pi(\cdot)$, we have $J_1(\pi(\cdot))=z^{(0)}$.
\end{corollary}

\proof
  (i) is immediately from the proof of the ``sufficient" part of Proposition \ref{Prop3.1}, and (ii) is from the proof of the ``necessary" part of Proposition \ref{Prop3.1}.   \rulex

\begin{remark}
  The condition \eqref{ftiao} is very mild. On the one hand, since \eqref{feasible} is a linear BSDE, its unique solution $\psi$ has following representation:
  \begin{equation*}
    \begin{aligned}
      \psi(t)=\mathbb{E}\left[(1-F(T)){\rm e}^{\int_t^T r(s)\mathrm{d}s }
      + \int_t^T f(s){\rm e}^{\int_t^s r(v)\mathrm{d}v }\mathrm{d}s |\mathcal{F}_t\right]
    \end{aligned}
  \end{equation*}
  (see Theorem 3.3 in Zhou and Yin \cite{Zhou2003} and Proposition 4.1 in Lim and Zhou \cite{Lim2002}).
   From Assumptions \ref{2.7}-\ref{2.9}, we know that $(1-F(T))\geq \varepsilon > 0$ and $f(s) \geq 0, s\in [0,T]$, and then $\psi(t) > 0$ for any $t\in[0,T]$. So \eqref{ftiao} is easily satisfied as long as there is one stock whose appreciation rate process is different from the interest rate process at any market mode, which is obviously a practically reasonable assumption. On the other hand, if \eqref{ftiao} fails, then Corollary \ref{Cor3.2}-(ii) implies that the mean-variance problem \eqref{question} is feasible only if $z=z^{(0)}$. This is a pathological and trivial case that does not warrant further consideration. So from this point, we shall assume that \eqref{ftiao} holds.
\end{remark}
After considering the feasibility, we proceed to study the issue of optimality. The mean variance problem with random horizon \eqref{question} is a dynamic optimization problem with a constraint $J_1(\pi(\cdot))=z$. In order to deal with the constraint, we employ the Lagrange multiplier technique. For any $\lambda\in \mathbb{R}$, define
\begin{equation}
  \begin{aligned}
    J(\pi(\cdot),\lambda)
    :&=J_{MV}(\pi(\cdot))+2\lambda\left(J_1(\pi(\cdot))-z\right)\\
    &=\mathbb{E}\left\{\int_0^Tf(t)\left[x(t)+(\lambda-z)\right]^2\mathrm{d}t + (1-F(T)) \left[x(T)+(\lambda-z)\right]^2\right\}-\lambda^2.
  \end{aligned}
\end{equation}
Our first goal is to solve the following unconstrained problem parameterized by the Lagrange multiplier $\lambda$:
\begin{equation}\label{unquestion}
  \begin{cases}
    {\rm minimize} \  J(\pi(\cdot),\lambda)\\
    {\rm subject \  to} \  (x(\cdot),\pi(\cdot)) \  {\rm admissible}.
  \end{cases}
\end{equation}
This is a stochastic LQ optimal control problem, which will be solved in the next two sections.

\section{Related Stochastic Riccati Equation}

When we study the unconstrained stochastic LQ problem \eqref{unquestion}, an equation known in the literature as the stochastic Riccati equation (SRE) will arise naturally. The existence and uniqueness of SRE play a fundamental role in the solution of the stochastic LQ problem \eqref{unquestion}. In this section, we study the solvability of the SRE and the auxiliary regime-switching BSDE.

We introduce the following regime-switching BSDEs:
\begin{equation}\label{SRE}
  \left\{
    \begin{aligned}
      &-\mathrm{d}p(t)=\Bigg[2f(t)+\left(2r(t)-\theta(t)\theta(t)^{\rm T}\right) p(t)-2\theta(t) {\it\Lambda}(t)- \frac{{\it\Lambda}(t)^{\rm T}{\it\Lambda}(t)}{p(t)} \Bigg] \mathrm{d}t \\
       &\qquad \qquad \quad- {\it\Lambda}(t)^{\rm T}\mathrm{d}W(t)-\hat{\it \Lambda}(t)^{\rm T}\mathrm{d}\tilde{\it \Phi}(t),\qquad t\in[0,T],\\
      &p(T)=2\big(1-F(T)\big),\\
      &p(t)> 0, \quad t\in[0,T].
    \end{aligned}
  \right.
\end{equation}

\begin{equation}\label{nonhom}
  \left\{
    \begin{aligned}
      &\mathrm{d}g(t)=\Bigg[\left(r(t)+\frac{2f(t)}{p(t)}\right)g(t)+\theta(t)\eta(t)-\frac{2f(t)}{p(t)} -\frac{\hat{\it\Lambda}(t)^{\rm T}\hat{\eta}(t)}{p(t)}\lambda(t) \Bigg]\mathrm{d}t\\
      &\qquad \qquad+\eta(t)^{\rm T}\mathrm{d}W(t)+\hat{\eta}(t)^{\rm T}\mathrm{d}\tilde{\it\Phi}(t),\quad t\in[0,T],\\
      &g(T)=1,
    \end{aligned}
  \right.
\end{equation}
where $\theta(t)=B(t)^{\rm T}\sigma(t)^{-1}$.

The regime-switching BSDE \eqref{SRE} is a special case of the SRE associated with the general stochastic LQ optimal control problems. And the auxiliary regime-switching BSDE \eqref{nonhom} will be used to handle the nonhomogeneous terms involved in the problem \eqref{unquestion}.
Now we introduce a subset of $\hat{S}_{\mathbb{F}}^{\infty}(0,T;\mathbb{R})$ as follows:
\begin{equation*}
  \begin{aligned}
    \hat{S}_{\mathbb{F}}^{\infty}(0,T;\mathbb{R}):=\big\{\varphi(\cdot)\in S_{\mathbb{F}}^{\infty}(0,T;\mathbb{R}) \ | \ {\rm there\ exist\ two\ real\ numbers}\ 0<b<B<\infty, \\
    {\rm such\  that}\ b\leq \varphi(t) \leq B\ {\rm for\ all}\ t\in [0,T] \big\}.
  \end{aligned}
\end{equation*}

The general SRE is a highly nonlinear, matrix-valued BSDE, and there are many results on its solvability, for example, Bismut \cite{Bismut1976}, Peng \cite{Peng1992}, Tang \cite{Tang2003}, Lim and Zhou \cite{Lim2002}, Yu \cite{Yu2013}, Shen et al. \cite{Shen2020} and the references therein. With the help of regime-switching BSDE of quadratic-exponential growth introduced by Shen et al. \cite{Shen2020}, we shall prove the existence and uniqueness of solution of SRE \eqref{SRE}.

Let us begin with another simpler regime-switching stochastic Riccati equation introduced by Shen et al. \cite{Shen2020} as follows:
\begin{equation}\label{fSRE}
  \left\{
    \begin{aligned}
      &-\mathrm{d}\bar{p}(t)=\left[\left(2r(t)-\theta(t)\theta(t)^{\rm T}\right)\bar{p}(t)-2\theta(t)\bar{\it \Lambda}(t) -\frac{\bar{\it \Lambda}(t)^{\rm T}\bar{\it \Lambda}(t)}{\bar{p}(t)}\right]\mathrm{d}t\\
      &\qquad\qquad\quad   -\bar{\it \Lambda}(t)^{\rm T}\mathrm{d}W(t) -\bar{\hat{\it \Lambda}}(t)^{\rm T}\mathrm{d}\tilde{\it \Phi}(t),\quad t\in [0,T],\\
      &\bar{p}(T)=2(1-F(T)),\\
      &\bar{p}(t)\geq 0,\quad t\in [0,T].
    \end{aligned}
  \right.
\end{equation}

\begin{lemma}[Shen et al. \cite{Shen2020}, Lemma 4.1]\label{4.1}
  The SRE \eqref{fSRE} admits a unique solution $\big(\bar{p}(\cdot),\bar{\it\Lambda}(\cdot),\\ \bar{\hat{\it\Lambda}} (\cdot)\big)   \in  \hat{S}_{\mathbb{F}}^ {\infty}(0,T;\mathbb{R})\times  \mathcal{H}^2_{{\rm BMO}_{\mathbf{P}}}
  (0,T;\mathbb{R}^n)\times \mathcal{J}^2_{{\rm BMO}_{\mathbf{P}}}(0,T;\mathbb{R}^N)$, satisfying {\rm (i)} $\frac{\bar{\it\Lambda}(\cdot)}{\bar{p}(\cdot)}\in \mathcal{H}^2_{{\rm BMO}_{\mathbf{P}}}(0,T;\mathbb{R}^n), \ \\ \frac{\bar{\hat{\it\Lambda}}(\cdot)}{\bar{p}(\cdot)}\in \mathcal{J}^2_{{\rm BMO}_{\mathbf{P}}} (0,T;\mathbb{R}^N)$;  {\rm (ii)} $\frac{\bar{\hat{\eta}}_k(\cdot)}{\bar{p}(\cdot)}\geq -1+\epsilon, \mathrm{d}t \otimes \mathrm{d}\mathbf{P} - a.e.$, for some $\epsilon >0$ and each $k=1,2,\cdots,N.$
\end{lemma}

\begin{theorem}\label{Thm4.2}
  Suppose the Assumptions \ref{2.7}-\ref{2.9} hold. Then there exists a unique solution $(p(\cdot),  {\it\Lambda}(\cdot), \hat{\it\Lambda}(\cdot))  \in \hat{S}_{\mathbb{F}}^ {\infty}(0,T;\mathbb{R})\times  \mathcal{H}^2_{{\rm BMO}_{\mathbf{P}}}   (0,T;\mathbb{R}^n)\times  \mathcal{J}^2_{{\rm BMO}_{\mathbf{P}}} (0,T;\mathbb{R}^N)$ for the stochastic Riccati equation \eqref{SRE}, satisfying {\rm (i)} $\frac{{\it\Lambda}(\cdot)}{{p}(\cdot)}\in \mathcal{H}^2_{{\rm BMO}_{\mathbf{P}}}(0,T;\mathbb{R}^n), \ \frac{{\hat{\it\Lambda}}(\cdot)}{{p}(\cdot)}\in \mathcal{J}^2_{{\rm BMO}_{\mathbf{P}}} (0,T;\mathbb{R}^N)$;  {\rm (ii)} $\frac{{\hat{\eta}}_k(\cdot)}{{p}(\cdot)}\geq -1+\epsilon, \mathrm{d}t \otimes \mathrm{d}\mathbf{P} - a.e.$, for some $\epsilon >0$ and each $k=1,2,\cdots,N.$
\end{theorem}
\proof
  From Lemma \ref{4.1}, the SRE \eqref{fSRE} admits a unique solution $(\bar{p}(\cdot),\bar{\it\Lambda}(\cdot), \bar{\hat{\it\Lambda}}(\cdot))\in  \hat{S}_{\mathbb{F}}^ {\infty}(0,T;\mathbb{R})\\ \times  \mathcal{H}^2_{{\rm BMO}_{\mathbf{P}}} (0,T;\mathbb{R}^n)\times \mathcal{J}^2_{{\rm BMO}_{\mathbf{P}}}(0,T;\mathbb{R}^N)$. Without loss of generality, we assume that $b\leq \bar{p}(t) \leq B$ for any $t\in [0,T]$ with two constants $0<b<B<\infty$. We consider the following BSDEs:
  \begin{equation}\label{ffSRE}
    \left\{
      \begin{aligned}
        &-\mathrm{d}\bar{p}(t)=\left[\left(2r(t)-\theta(t)\theta(t)^{\rm T}\right)\bar{p}(t)-2\theta(t)\bar{\it\Lambda}(t) -\frac{\bar{\it\Lambda}(t)^{\rm T}\bar{\it\Lambda}(t)}{\bar{p}(t)\vee b}\right]\mathrm{d}t\\
        &\qquad\qquad\quad   -\bar{\it\Lambda}(t)^{\rm T}\mathrm{d}W(t) -\bar{\hat{\it\Lambda}}(t)^{\rm T}\mathrm{d}\tilde{\it\Phi}(t),\quad t\in [0,T],\\
        &\bar{p}(T)=2(1-F(T)),\\
        &\bar{p}(t)\geq 0,\quad t\in [0,T],
      \end{aligned}
    \right.
  \end{equation}
  and
  \begin{equation}\label{fffSRE}
    \left\{
      \begin{aligned}
        &-\mathrm{d}{p}(t)=\left[2f(t)+\left(2r(t)-\theta(t)\theta(t)^{\rm T}\right){p}(t)-2\theta(t){\it\Lambda}(t) -\frac{{\it\Lambda}(t)^{\rm T}{\it\Lambda}(t)}{{p}(t)\vee b}\right]\mathrm{d}t\\
        &\qquad\qquad\quad   -{\it\Lambda}(t)^{\rm T}\mathrm{d}W(t) -{\hat{\it\Lambda}}(t)^{\rm T}\mathrm{d}\tilde{\it\Phi}(t),\quad t\in [0,T],\\
        &{p}(T)=2(1-F(T)),\\
        &{p}(t)> 0,\quad t\in [0,T].
      \end{aligned}
    \right.
  \end{equation}
  Obviously, the unique solution $(\bar{p}(\cdot),\bar{\it\Lambda}(\cdot),\bar{\hat{\it\Lambda}}(\cdot))$ of SRE \eqref{fSRE} also hold for BSDE \eqref{ffSRE}. Since $f(\cdot)$, $r(\cdot)$, and $\theta(\cdot)$ are bounded and nonnegative, the driver $h(t,p,{\it\Lambda},\hat{\it\Lambda})=2f(t)+(2r(t)-\theta(t)
  \theta(t)^{\rm T})p-2\theta(t){\it\Lambda}-\frac{{\it\Lambda}^{\rm T}{\it\Lambda}}{p\vee b}$ of BSDE \eqref{fffSRE} satisfies
  \begin{equation*}
    |h(t,p,{\it\Lambda},\hat{\it\Lambda})| \leq C(1+|p|+|{\it\Lambda}|^2),
  \end{equation*}
  where $C$ is a constant. Therefore, from Lemma A.1 in Shen et al. \cite{Shen2020}, there exists a unique bounded solution $(p^{(b)}(\cdot),{\it\Lambda}^{(b)}(\cdot),\hat{\it\Lambda}^{(b)}(\cdot)) \in  {S}_{\mathbb{F}}^ {\infty}(0,T;\mathbb{R})\times  \mathcal{H}^2_{{\rm BMO}_{\mathbf{P}}} (0,T;\mathbb{R}^n)\times \mathcal{J}^2_{{\rm BMO}_{\mathbf{P}}}(0,T;\mathbb{R}^N)$ for BSDE \eqref{fffSRE}. Next we apply the comparison theorem and from
  \begin{equation*}
    h(t,p,{\it\Lambda},\hat{\it\Lambda}) \geq (2r(t)-\theta(t)\theta(t)^{\rm T})p-2\theta(t){\it\Lambda}-\frac{{\it\Lambda}^{\rm T}{\it \Lambda}} {p\vee b},
  \end{equation*}
  we deduce that $p^{(b)}(t)\geq \bar{p}(t)\geq b$. Hence $(p^{(b)}(\cdot),{\it\Lambda}^{(b)}(\cdot),\hat{\it\Lambda}^{(b)}(\cdot)) \in  \hat{S}_{\mathbb{F}}^ {\infty}(0,T;\mathbb{R})\times  \mathcal{H}^2_{{\rm BMO}_{\mathbf{P}}} (0,T;\mathbb{R}^n)\\
  \times \mathcal{J}^2_{{\rm BMO}_{\mathbf{P}}}(0,T;\mathbb{R}^N)$ is the unique solution of SRE \eqref{SRE}.
  The proof of the solution's properties are similar to the proof in Appendix of Shen et al. \cite{Shen2020}, so we omit it.  \rulex

Then we show that the existence and uniqueness of solution of \eqref{nonhom}.

\begin{theorem}
  Suppose the Assumptions \ref{2.7}-\ref{2.9} hold. Then there exists a unique solution $\big(g(\cdot),
  \eta(\cdot),\hat{\eta}(\cdot)\big)\in S_{\mathbb{F}}^{2}(0,T;\mathbb{R})\times L_{\mathbb{F}}^2(0,T;\mathbb{R}^n)\times {\it\Pi}^2_{\mathbb{F}}(0,T;\mathbb{R}^N)$ for the BSDE \eqref{nonhom}. Moreover, we have $0<g(t)\leq 1$ for any $t\in [0,T]$ and if $r(t)>0,\ a.e.\ t\in [0,T]$, then $0<g(t)<1$ for all $t\in [0,T)$.
\end{theorem}

\proof
  Since all the market parameters are bounded, the stochastic integral $-\int_0^t\theta(s)\mathrm{d}W (s)$ is a $\text{BMO}_{\textbf{P}}$ martingale. Then we define a probability measure $\mathbf{Q}_1$ equivalent to $\mathbf{P}$ as \begin{equation}\label{Q1}
    \left.\frac{\mathrm{d}\mathbf{Q}_1}{\mathrm{d}\mathbf{P}}\right|_{\mathcal{F}_T}=\Theta_1:=\mathcal{E}\left(-\int_0^T \theta(s)\mathrm{d}W(s)\right).
  \end{equation}
  where the stochastic exponential $\mathcal{E}(M)$ is given by $\mathcal{E}(M)={\rm e}^{(M-\frac{1}{2}\langle M \rangle)}$. Under $\mathbf{Q}_1$, by Girsanov's theorem, the process
  \begin{equation}
    W^{\mathbf{Q}_1}(t):=W(t)+\int_0^t \theta(s)\mathrm{d}s
  \end{equation}
  is a $n$-dimensional standard Brownian motion, and $\tilde{\it\Phi}^{\mathbf{Q}_1}(t):=\tilde{\it\Phi}(t)$ has the same probability law as under $\mathbf{P}$ which is still an $\mathbb{R}^N$-valued jump martingale associated with the chain.

  Since $\frac{\hat{\it\Lambda}}{p}\in \mathcal{J}^2_{{\rm BMO}_{\mathbf{P}}}(0,T;\mathbb{R}^N)$ and the probability measure $\mathbf{Q}_1$ is constructed as \eqref{Q1}, it can be shown as Theorem 3.3 in \cite{Kazamaki2006} that $\frac{\hat{\it\Lambda}}{p}\in \mathcal{J}^2_{{\rm BMO}_{\mathbf{Q}_1}} (0,T;\mathbb{R}^N)$. By the equivalence of $\mathbf{Q}_1$ and $\mathbf{P}$, we have that $\frac{{\hat{\eta}}_k(\cdot)}{{p}(\cdot)}\geq -1+\epsilon,\ \mathrm{d}t \otimes \mathrm{d}\mathbf{Q}_1-a.e.$. Hence, the techniques of \cite{Kazamaki2006} and the jump sizes ${\it\Delta \Phi}(t)$ can be used to show that $\mathcal{E}\left(\int_0^t \frac{\hat{\it\Lambda}(s)^{\rm T}}{p(s)}\mathrm{d}\tilde{\it\Phi} ^{\mathbf{Q}_1}(s)\right)$ is a uniformly integrable $(\mathbb{F}, \mathbf{Q}_1)$-martingale. Therefore, we can define a probability measure $\mathbf{Q}_2$ equivalent to $\mathbf{Q}_1$ as
  \begin{equation}
    \left.\frac{\mathrm{d}\mathbf{Q}_2}{\mathrm{d}\mathbf{Q}_1}\right|_{\mathcal{F}_T}=\Theta_2:=\mathcal{E}\left(\int_0^T \frac{\hat{\it\Lambda}(s)^{\rm T}}{p(s)}\mathrm{d}\tilde{\it\Phi}^{\mathbf{Q}_1}(s)\right).
  \end{equation}
  Under $\mathbf{Q}_2$, the $\mathbf{Q}_1$-Brownian motion $W^{\mathbf{Q}_1}$ is still a $n$-dimensional standard Brownian motion, which is denoted by $W^{\mathbf{Q}_2}(t):=W^{\mathbf{Q}_1}(t)$. And the process $\tilde{\it\Phi}^{\mathbf{Q}_2}(t):=\tilde{\it\Phi}^{\mathbf{Q}_1}(t)-\int_0^t\frac{\hat{\it\Lambda}(s)}{p(s)}\lambda(s) \mathrm{d}s$ is an $\mathbb{R}^{N}$-valued martingale associated with the chain $\alpha$. So the $\mathbf{Q}_2$-intensity of ${\it\Phi}$ is given by an $\mathbb{R}^N$-valued process:
  \begin{equation}
    \lambda^{\mathbf{Q}_2}(t):=\left[\lambda_1(t)\left(1+\frac{\hat{\it\Lambda}_{1}(t)}{p(t)}\right), \lambda_2(t)\left(1+\frac{\hat{\it\Lambda}_{2}(t)}{p(t)}\right),\dots, \lambda_N(t)\left(1+\frac{\hat{\it\Lambda}_{N}(t)}{p(t)}\right)\right]^{\rm T}.
  \end{equation}
  Therefore, the linear BSDE \eqref{nonhom} under $\mathbf{Q}_2$ becomes
  \begin{equation}\label{QBSDE}
    \left\{
      \begin{aligned}
        &\mathrm{d}g(t)=\left[-\frac{2f(t)}{p(t)}+\left(r(t)+\frac{2f(t)}{p(t)}\right)g(t)\right]\mathrm{d}t +\eta(t)^{\rm T}\mathrm{d}W^{\mathbf{Q}_2}(t)+\hat{\eta}(t)^{\rm T}\mathrm{d}\tilde{\it\Phi}^{\mathbf{Q}_2}(t),\\
        &g(T)=1.
      \end{aligned}
    \right.
  \end{equation}
  It is easy to obtain the well-posedness of \eqref{QBSDE} which admits a unique square integrable solution $(g(\cdot),\eta(\cdot),\hat{\eta}(\cdot))$ under the probability measure $\mathbf{Q}_2$. Moreover, since $f(t)$ and $p(t)$ are bounded and $p(t)\geq b>0$, we have
  \begin{equation*}
    g(t)=\mathbb{E}^{\mathbf{Q}_2}\left[{\rm e}^{-\int_t^T\left(r(s)+\frac{2f(s)}{p(s)}\right) \mathrm{d}s }+\int_t^T \frac{2f(s)}{p(s)}{\rm e}^{ -\int_t^s \left(r(v)+\frac{2f(v)}{p(v)}\right)\mathrm{d}v}\mathrm{d}s|\mathcal{F}_t\right].
  \end{equation*}
  It is easy to see that $g(\cdot)$ is a bounded process and $g(t)>0$ for any $t\in [0,T]$. Next we introduce another BSDE:
  \begin{equation}\label{ANBSDE}
    \left\{
      \begin{aligned}
        &\mathrm{d}\tilde{g}(t)=\left[-\frac{2f(t)}{p(t)}+\frac{2f(t)}{p(t)}\tilde{g}(t)\right]\mathrm{d}t +\tilde{\eta}(t)^{\rm T}\mathrm{d}W^{\mathbf{Q}_2}(t)+\tilde{\hat{\eta}}(t)^{\rm T} \mathrm{d} \tilde{\it\Phi}^{\mathbf{Q}_2}(t),\\
        &\tilde{g}(T)=1.
      \end{aligned}
    \right.
  \end{equation}
  Obviously, the unique solution is  $(\tilde{g}(\cdot),\tilde{\eta}(\cdot),\tilde{\hat{\eta}}(\cdot))\equiv(1,0,0)$  to \eqref{ANBSDE}. Denoting $\bar{g}(\cdot)=\tilde{g}(\cdot)-g(\cdot)$, $\bar{\eta}(\cdot)=\tilde{\eta}(\cdot)-\eta(\cdot)$ and $\bar{\hat{\eta}}(\cdot)=\tilde{\hat{\eta}}(\cdot)-\hat{\eta}(\cdot)$, then $(\bar{g}(\cdot), \bar{\eta}(\cdot), \bar{\hat{\eta}}(\cdot))$ satisfies the following BSDE:

  \begin{equation}
    \left\{
      \begin{aligned}
        &\mathrm{d}\bar{g}(t)=\left[-r(t)g(t)+\frac{2f(t)}{p(t)}\bar{g}(t)\right]\mathrm{d}t +\bar{\eta}(t)^{\rm T}\mathrm{d}W^{\mathbf{Q}_2}(t)+\bar{\hat{\eta}}(t)^{\rm T} \mathrm{d} \tilde{\it\Phi}^{\mathbf{Q}_2}(t),\\
        &\bar{g}(T)=0.
      \end{aligned}
    \right.
  \end{equation}
  Explicitly, $\bar{g}(t)=\mathbb{E}^{\mathbf{Q}_2}\left[\int_t^T r(s)g(s){\rm e}^{-\int_t^s\frac{2f(v)}{p(v)} \mathrm{d}v} \mathrm{d}s\right]$. Due to $r(t)\geq 0$ and $g(t)>0$, we obtain $\bar{g}(t)\geq 0$. From the definition of $\bar{g}(t)=\tilde{g}(t)-g(t)$, we get $g(t)\leq 1,\ t\in [0,T]$. Moreover, we have $g(t)<1$ for any $t\in[0,T)$, if the interest rate $r(t)>0,\ a.e.\ t\in [0,T]$.  \rulex

\section{Solution to the Unconstrained Problem}

Now we give the solution of the unconstrained problem \eqref{unquestion}.

\begin{theorem}\label{Thm5.1}
  Under Assumptions \ref{2.7}-\ref{2.9}, the problem \eqref{unquestion} is solvable. The unique optimal feedback control for $t\in[0,T]$ is given by
  \begin{equation}\label{optimalcontrol}
    \pi^{(\lambda)}(t)=-\sigma(t)^{-1}\left(\theta(t)+\frac{{\it\Lambda}(t)}{p(t-)}\right)\left[ x^{(\lambda)}(t-) + (\lambda-z)g(t-) \right]-(\lambda-z)\sigma(t)^{-1}\eta(t).
  \end{equation}
  The corresponding optimal state trajectory is given by
  \begin{equation}\label{SSSDE}
    \begin{aligned}
      \mathrm{d}x^{(\lambda)}(t)&=\Bigg\{r(t)x^{(\lambda)}(t)-\theta(t)\left(\theta(t)+\frac{{\it\Lambda}(t)} {p(t)}\right)\left[x^{(\lambda)}(t) +(\lambda-z)g(t)\right]-(\lambda-z)\theta(t)\eta(t)\Bigg\}\mathrm{d}t\\
      &\quad\quad-\Bigg\{\left(\theta(t)^{\rm T}+\frac{{\it\Lambda}(t)^{\rm T}} {p(t)}\right)[x^{(\lambda)}(t)+(\lambda-z)g(t)] -(\lambda-z)\eta(t)^{\rm T}\Bigg\}\mathrm{d}W(t).
    \end{aligned}
  \end{equation}
  And the associated optimal cost is given by
  \begin{equation}\label{optimalcost}
    \begin{aligned}
      J(\pi^{(\lambda)}(\cdot),\lambda)
      =\left[\frac{1}{2}p(0)g(0)^2+{\it\Delta}-1\right](\lambda-z)^2  +\left[p(0)g(0)x_0-2z\right](\lambda-z)+\frac{1}{2}p(0)x_0^2-z^2,
    \end{aligned}
  \end{equation}
  where
  \begin{equation}\label{Delta}
    {\it\Delta}:=\mathbb{E}\left[\int_0^T f(t)|g(t)-1|^2+\frac{1}{2}p(t)\hat{\eta}(t)^2\lambda(t) \mathrm{d}t\right]\geq 0.
  \end{equation}
\end{theorem}

Before proving this theorem, we verify that the strategy $\pi^{(\lambda)}(\cdot)$ is admissible. 

\begin{lemma}\label{5.2}
  Let Assumptions \ref{2.7}-\ref{2.9} hold. $\pi^{(\lambda)}(\cdot)$ defined by \eqref{optimalcontrol} is admissible.
\end{lemma}
\proof
  Substituting the feedback control \eqref{optimalcontrol} into the wealth equation \eqref{wealth}, we have \eqref{SSSDE}. In \eqref{SSSDE}, the term $\Lambda(t)$ associated with the SRE \eqref{SRE} appears in the coefficients of the variable $x$ in both drift and diffusions. Although linear, the issue of existence and uniqueness of solution of \eqref{SSSDE} no longer lies in the domain of standard theory, which requires the coefficients of $x$ to be uniformly bounded. However, from Lemma 4.3 in Shen et al. \cite{Shen2020}, we directly get the desired results that the SDE \eqref{SSSDE} admits a unique strong solution. In addition, the left-limit processes $x^{(\lambda)}(t-)$, $p(t-)$ and $g(t-)$ are predictable  as $x^{(\lambda)}(t)$, $p(t)$ and $g(t)$ are $\mathbb{F}$-adapted, ${\rm c\grave{a}dl\grave{a}g}$ processes. As a consequence that $\Lambda(t)$ and $\eta(t)$ are also predictable, we can deduce that the strategy $\pi^{(\lambda)}$ is predictable. Next we want to verify that $\pi^{(\lambda)}$ is square integrable.

  Applying It\^o's formula to $p(t)(x^{(\lambda)}(t)+(\lambda-z)g(t))^2$, we have
  \begin{equation}\label{ItoBSDE}
    \begin{aligned}
      &\mathrm{d}\left\{p(t)\left[x^{(\lambda)}(t)+(\lambda-z)g(t)\right]^2\right\}\\
      &=\Bigg\{2p(t)\left[x^{(\lambda)}(t)+(\lambda-z)g(t)\right]\bigg\{r(t)x^{(\lambda)}(t)+B(t)\pi^{(\lambda)} (t)+(\lambda-z) \bigg[r(t)g(t)
      +2\frac{f(t)}{p(t)}(g(t)-1)\\
      &\quad\quad+\theta(t)\eta(t) -\frac{\hat{\it\Lambda}(t)^{\rm T}\hat{\eta}(t)}{p(t)}\lambda(t) \big)\bigg]\bigg\}
      +p(t)\Big[\left(\sigma(t)\pi^{(\lambda)}(t))+(\lambda-z)\eta(t)\right)^2+(\lambda-z)^2\hat{\eta}(t)^2\lambda(t)\Big]\\   &\quad\quad-\left[x^{(\lambda)}(t)+(\lambda-z)g(t)\right]^2 \bigg[2f(t)+\big[2r(t)-\theta(t)\theta(t)^{\rm T}\big]
      p(t)-2\theta(t){\it\Lambda}(t)-\frac{{\it\Lambda}(t)^{\rm T}{\it\Lambda}(t)} {p(t)}\bigg]\\
      &\quad\quad+2[x^{(\lambda)}(t)+(\lambda-z)g(t)]\left({\it\Lambda}(t)\left[\sigma(t)\pi^{(\lambda)}(t)+(\lambda-z)\eta(t)\right]
      +(\lambda-z) \lambda(t)\hat{ \it\Lambda}(t)\hat{\eta}(t)\right) \Bigg\}\mathrm{d}t\\
      &\quad\quad +N^{(\lambda)}(t)\mathrm{d}W(t)+\tilde{N}^{(\lambda)}(t)\mathrm{d}\tilde{\it\Phi}(t)\\
      &=\bigg\{p(t)\bigg[\sigma(t)\pi^{(\lambda)}(t)+(\lambda-z)\eta(t)+ \left(\theta(t)) +\frac{{\it\Lambda}(t)}{p(t)} \right)(x^{(\lambda)}(t)+(\lambda-z)g(t))\bigg]^2\\
      &\quad\quad+2(\lambda-z)^2f(t)(g(t)-1)^2
      +(\lambda-z)^2p(t) \hat{\eta}(t)^2\lambda(t)-2f(t)\left[x^{(\lambda)}(t)+(\lambda-z)\right]^2\bigg\}\mathrm{d}t \\
      &\quad\quad +N^{(\lambda)}(t)\mathrm{d}W(t)+\tilde{N}^{(\lambda)}(t)\mathrm{d}\tilde{\it\Phi}(t)\\
      &=\bigg\{ 2(\lambda-z)^2f(t)(g(t)-1)^2
      +(\lambda-z)^2p(t) \hat{\eta}(t)^2\lambda(t)-2f(t)\left[x^{(\lambda)}(t)+(\lambda-z)\right]^2\bigg\}\mathrm{d}t \\
      &\quad\quad +N^{(\lambda)}(t)\mathrm{d}W(t)+\tilde{N}^{(\lambda)}(t)\mathrm{d}\tilde{\it\Phi}(t)
    \end{aligned}
  \end{equation}
  where
  \begin{equation*}
    \begin{aligned}
      N^{(\lambda)}(t)=2p(t)\left[x^{(\lambda)}(t)+(\lambda-z)g(t)\right]\left[\pi^{(\lambda)}(t)\sigma(t)+(\lambda-z)\eta(t)\right] +{\it\Lambda}(t)\left[x^{(\lambda)}(t)+(\lambda-z)g(t)\right]^2,\\
    \end{aligned}
  \end{equation*}
  \begin{equation*}
    \tilde{N}^{(\lambda)}(t)=2(\lambda-z)p(t-)\hat{\eta}(t)\left[x^{(\lambda)}(t-)+(\lambda-z)g(t-)\right] +\hat{\it\Lambda}(t)\left[x^{(\lambda)}(t-)+(\lambda-z)g(t-)\right]^2.
  \end{equation*}
  For any given $t\in [0,T]$ and any given $\mathbb{F}$-stopping time $\tau$, we integrate \eqref{ItoBSDE} on $[0,t\wedge\tau]$ and get
  \begin{equation*}
    \begin{aligned}
      &p(t\wedge \tau)\left[x^{(\lambda)}(t\wedge \tau)+(\lambda-z)g(t\wedge \tau)\right]^2-p(0)\left[x^{(\lambda)}(0)+(\lambda-z)g(0)\right]^2\\
      &=\int_0^{t\wedge \tau}\bigg\{ 2(\lambda-z)^2f(s)(g(s)-1)^2
      +(\lambda-z)^2p(s) \hat{\eta}(s)^2\lambda(s)-2f(s)\left[x^{(\lambda)}(s)+(\lambda-z)\right]^2\bigg\}\mathrm{d}s\\
      &\quad\quad +\int_0^{t\wedge \tau}N^{(\lambda)}\mathrm{d}W(s)+\int_0^{t\wedge \tau}\tilde{N}^{(\lambda)}\mathrm{d}\tilde{\it\Phi}(s).
    \end{aligned}
  \end{equation*}
  Since $\int_0^{t\wedge \tau} N^{(\lambda)}\mathrm{d}s$ and $\int_0^{t\wedge \tau} \tilde{N}^{(\lambda)} \mathrm{d}\tilde{\it\Phi}(s)$ are locally square integrable martingales, there exists a sequence of stopping times $\{\tau_i\}_{i=1}^{\infty}$ which increases and diverges $\mathbf{P}-a.s.$, as $i\rightarrow \infty$. It follows that
  \begin{equation}\label{lemma1}
    \begin{aligned}
      \mathbb{E}&\left[p(t\wedge \tau \wedge \tau_i)\left(x^{(\lambda)}(t\wedge \tau \wedge \tau_i)+(\lambda-z)g(t\wedge \tau \wedge \tau_i)\right)^2\right]\\
      &\leq p(0)\left[x^{(\lambda)}(0)+(\lambda-z)g(0)\right]^2
      +\mathbb{E}\int_0^{t\wedge \tau\wedge \tau_i}(\lambda-z)^2\bigg\{ 2f(s)(g(s)-1)^2+p(s) \hat{\eta}(s)^2\lambda(s) \bigg\}\mathrm{d}s.
    \end{aligned}
  \end{equation}
  Owing to $0<b\leq p(s)\leq B$, $f(\cdot),\ g(\cdot)$ are non-negative bounded and $\hat{\eta}(\cdot) \in {\it\Pi}^2_{\mathbb{F}} (0,T;\mathbb{R}^N)$, then it is easy to see that the integrand in the right-hand side of \eqref{lemma1} is non-negative. And we also know that the left-hand side is dominated by an integrable random variable $B\sup_{t\in [0,T]} |x(t)+(\lambda-z)g(t)|^2$. Therefore, by sending $i$ to $\infty$ and applying the dominated convergence theorem and the monotone convergence theorem to \eqref{lemma1}, we obtain
  \begin{equation*}
    \begin{aligned}
      \mathbb{E}&\left[p(t\wedge \tau)\left(x^{(\lambda)}(t\wedge \tau)+(\lambda-z)g(t\wedge \tau )\right)^2\right]\\
      &\quad\quad\leq p(0)\left[x^{(\lambda)}(0)+(\lambda-z)g(0)\right]^2
      +\mathbb{E}\int_0^{t\wedge \tau}(\lambda-z)^2\bigg\{ 2f(s)(g(s)-1)^2+p(s) \hat{\eta}(s)^2\lambda(s) \bigg\}\mathrm{d}s.
    \end{aligned}
  \end{equation*}
  Then we have
  \begin{equation*}
    \mathbb{E}\left[\left(x^{(\lambda)}(t\wedge \tau)+(\lambda-z)g(t\wedge \tau)\right)^2\right]< \infty.
  \end{equation*}
  Moreover, we can get
  \begin{equation*}
    \begin{aligned}
      \mathbb{E}\left[\left|x^{(\lambda)}(t\wedge\tau)\right|^2\right]
      &=\mathbb{E}\left[\left|\left(x^{(\lambda)}(t\wedge \tau)+(\lambda-z)g(t\wedge \tau)\right) -(\lambda-z)g(t\wedge \tau)\right|^2\right]\\
      &\leq 2\mathbb{E}\left[\left|x^{(\lambda)}(t\wedge \tau)+(\lambda-z)g(t\wedge \tau)\right|^2\right] +2\mathbb{E}\left[\left|(\lambda-z)g(t\wedge \tau)\right|^2\right].
    \end{aligned}
  \end{equation*}
  That is, for any $t\in [0,T]$ and any stopping time $\tau$, we obtain
  \begin{equation}\label{jiben}
    \mathbb{E}\left[\left|x^{(\lambda)}(t\wedge \tau)\right|^2\right]< \infty.
  \end{equation}
  Now, we show that \eqref{jiben} implies the admissibility of $\pi^{(\lambda)}(\cdot)$. Applying It\^o's formula, we have
  \begin{equation*}
    \begin{aligned}
      |x^{(\lambda)}(t)|^2=x_0^2&+\int_0^t 2x^{(\lambda)}(s)\left[r(s)x^{(\lambda)}(s)+\pi^{(\lambda)}(s)^{\rm T}B(s)\right]\mathrm{d}s \\
      & +\int_0^t|\pi^{(\lambda)}(s)^{\rm T}\sigma(s)|^2\mathrm{d}s
      +\int_0^t 2x^{(\lambda)}(s)\pi^{(\lambda)}(s)\sigma(s)\mathrm{d}W(s).
    \end{aligned}
  \end{equation*}
  From the inequality $-2xB\cdot \pi=-2(\sqrt{2}x\theta)\cdot(\pi \sigma /\sqrt{2})\leq 2x^2|\theta|^2+|\pi\sigma|^2/2$, we have
  \begin{equation*}
    \begin{aligned}
      |x^{(\lambda)}(t)|^2\geq x_0^2&-\int_0^t 2|x^{(\lambda)}(s)|^2\left[|\theta(s)|^2-r(s)\right]\mathrm{d}s \\ &+\frac{1}{2}\int_0^t|\pi^{(\lambda)}(s)^{\rm T}\sigma(s)|^2\mathrm{d}s
      +\int_0^t 2x^{(\lambda)}(s)\pi^{(\lambda)}(s)\sigma(s)\mathrm{d}W(s).
    \end{aligned}
  \end{equation*}
  Note that $2x^{(\lambda)}(\cdot)\pi^{(\lambda)}(\cdot)\sigma(\cdot)\in L^{2,loc}_{\mathbb{F}} (0,T;\mathbb{R}^n).$ Therefore, there exists a localizing sequence of stopping times $\{\sigma_i\}_{i=1}^{\infty}$ which increases and diverges $\mathbf{P}-a.s.$, such that
  \begin{equation*}
    x_0^2+\frac{1}{2}\mathbb{E}\int_0^{T\wedge \sigma_i}|\pi^{(\lambda)}(t)\sigma(t)|^2\mathrm{d}t \leq \mathbb{E}\left[|x^{(\lambda)}(T\wedge \sigma_i)|^2\right]+2\mathbb{E}\int_0^{T\wedge \sigma_i}|x^{(\lambda)}(s)|^2\left[|\theta(s)|^2-r(s)\right]\mathrm{d}s.
  \end{equation*}
  By virtue of \eqref{jiben} and Assumption \ref{2.1}, we have
  $$\mathbb{E}\int_0^{T\wedge \sigma_i}|\pi^{(\lambda)}(t)|^2\mathrm{d}t<\infty.$$
  Sending $i$ to $\infty$, we can directly get
  $$\mathbb{E}\int_0^{T}|\pi^{(\lambda)}(t)|^2\mathrm{d}t<\infty.$$
  Then we get the admissibility of $\pi^{(\lambda)}(\cdot)$. Consequently, the corresponding wealth process $x^{(\lambda)}(\cdot) \in S^2_{\mathbb{F}}(0,T;\mathbb{R})$.   \rulex

Then we prove Theorem \ref{Thm5.1}.

\proof
  Let $\pi(\cdot)\in\mathcal{U}$ be any given admissible control and $x(\cdot)$ be the corresponding state trajectory. Similar to Lemma \ref{5.2}, applying It\^o's formula to $p(t)\left(x(t)+(\lambda-z)g(t)\right)^2$, we obtain
  \begin{equation*}
    \begin{aligned}
      p(t)&\left[x(t)+(\lambda-z)g(t)\right]^2+\int_0^t 2f(s)\left[x(s)+(\lambda-z)\right]^2\mathrm{d}s\\
      &=\int_0^t \bigg\{p(s)\bigg[\sigma(s)\pi(s)+(\lambda-z)\eta(s)+ \left(\theta(s) +\frac{\Lambda(s)}{p(s)} \right)(x(s)+(\lambda-z)g(s))\bigg]^2\\
      &\quad\quad+2(\lambda-z)^2f(s)(g(s)-1)^2
      +(\lambda-z)^2p(s) \hat{\eta}(s)^2\lambda(s)\bigg\}\mathrm{d}s+p(0)\left(x(0)+(\lambda-z)g(0)\right)^2 \\
      &\quad\quad +\int_0^t N(s)\mathrm{d}W(s)
      +\int_0^t \tilde{N}(s)\mathrm{d}\tilde{\it\Phi}(s)
    \end{aligned}
  \end{equation*}
  where
  \begin{equation*}
    \begin{aligned}
      N(t)&=2p(s)\left[x(t)+(\lambda-z)g(t)\right]\left(\pi(t)\sigma(t)+(\lambda-z)\eta(t)\right) +{\it\Lambda}(t)\left[x(t)+(\lambda-z)g(t)\right]^2 \\
      \tilde{N}(t)&=2(\lambda-z)p(t-)\hat{\eta}(t)\left[x(t-)+(\lambda-z)g(t-)\right] +\hat{\it\Lambda}(t)\left[x(t-)+(\lambda-z)g(t-)\right]^2
    \end{aligned}
  \end{equation*}
  Similar to Lemma \ref{5.2}, integrating the above from $0$ to $T$ and taking expectations, we obtain
  \begin{equation*}
    \begin{aligned}
      &\mathbb{E}\big[2(1-F(T))\left(x(T)+(\lambda-z)\right)^2\big]+\mathbb{E}\left[2\int_0^Tf(t)\left( x(t)+(\lambda-z) \right)^2\mathrm{d}t\right]\\
      &=\mathbb{E}\int_0^T p(t)\left|\sigma(t)\pi(t)+\left(\theta(t)+\frac{{\it\Lambda}(t)}{p(t-)}\right)\left[ x(t-) + (\lambda-z)g(t-) \right]+(\lambda-z)\eta(t)\right|^2\mathrm{d}t\\
      &\quad+p(0)\left[x_0+(\lambda-z)g(0)\right]^2+2(\lambda-z)^2\mathbb{E}\int_0^T f(t)\left(g(t)-1\right)^2\mathrm{d}t +(\lambda-z)^2\mathbb{E}\int_0^T p(t) \hat{\eta}(t)^2\lambda(t)\mathrm{d}t\\
    \end{aligned}
  \end{equation*}
  Consequently,
  \begin{equation*}
    \begin{aligned}
      J(\pi^{(\lambda)}(\cdot),\lambda)&=\mathbb{E}\left[\int_0^Tf(t)[x(t)+(\lambda-z)]^2\mathrm{d}t+(1-F(T)) [x(T)+(\lambda-z)]^2\right]-\lambda^2\\
      &=\frac{1}{2}\mathbb{E}\int_0^T p(t)\left|\sigma(t)\pi(t)+\left(\theta(t)+\frac{{\it\Lambda}(t)}{p(t-)}\right)\left[ x(t-) + (\lambda-z)g(t-) \right]+(\lambda-z)\eta(t)\right|^2\mathrm{d}t\\
      &\quad\quad+\frac{1}{2}p(0)\left[x_0+(\lambda-z)g(0)\right]^2+(\lambda-z)^2\mathbb{E}\int_0^T f(t)\left(g(t)-1\right)^2\mathrm{d}t\\
     &\quad\quad+\frac{1}{2}(\lambda-z)^2\mathbb{E}\int_0^T p(t) \hat{\eta}(t)^2\lambda(t)\mathrm{d}t-\lambda^2.
    \end{aligned}
  \end{equation*}
  Since $p(t)>0$ by Theorem \ref{Thm4.2}, it follows immediately that the optimal feedback control is given by \eqref{optimalcontrol} and the optimal value is given by
  \begin{equation*}
    \begin{aligned}
      J(\pi^{(\lambda)}(\cdot),\lambda)
      &=\frac{1}{2}p(0)\left[x_0+(\lambda-z)g(0)\right]^2+(\lambda-z)^2\mathbb{E}\int_0^T f(t)\left(g(t)-1\right)^2\mathrm{d}t\\
      &\quad\quad +\frac{1}{2}(\lambda-z)^2\mathbb{E}\int_0^T p(t) \hat{\eta}(t)^2\lambda(t)\mathrm{d}t-\lambda^2\\
      &=\left(\frac{1}{2}p(0)g(0)^2+{\it\Delta}-1\right)(\lambda-z)^2  +\left[p(0)g(0)x_0-2z\right](\lambda-z)+\frac{1}{2}p(0)x_0^2-z^2,
    \end{aligned}
  \end{equation*}
  where ${\it\Delta}$ is defined by \eqref{Delta}.   \rulex

\section{Efficient frontier}

In this section, proceed to derive the efficient frontier for the original mean-variance problem \eqref{question}.

\begin{theorem}
  Let Assumptions \ref{2.7}-\ref{2.9} and \eqref{ftiao} hold. Then
  \begin{equation}\label{quadratic}
    \frac{1}{2}p(0)g(0)^2+{\it\Delta}-1<0.
  \end{equation}
  Moreover, the efficient portfolio corresponding to $z$ is
  \begin{equation}\label{opcontrol}
    \pi^*(t):=\pi^{(\lambda^*)}(t)=-\left(\theta(t)+\frac{{\it\Lambda}(t)}{p(t-)}\right)\sigma(t)^{-1} \left[x(t-)+(\lambda^*-z)g(t-) \right]-(\lambda^*-z)\sigma(t)^{-1}\eta(t)^{\rm T},
  \end{equation}
  where
  \begin{equation}\label{optimallambda}
    \lambda^*-z=\frac{2z-p(0)g(0)x_0}{p(0)g(0)^2+2{\it\Delta}-2}.
  \end{equation}
  Furthermore, the optimal value of ${\rm Var} x(\tau\wedge T)$, among all the wealth processes $x(\cdot)$ satisfying $\mathbb{E}x(\tau\wedge T)=z$, is
  \begin{equation}\label{optimalvalue}
    \begin{aligned}
      {\rm Var } x^*(\tau\wedge T)=\frac{2{\it\Delta}+p(0)g(0)^2}{2-2{\it\Delta}-p(0) g(0)^2} \left(z-\frac{p(0) g(0)}{2{\it\Delta}+p(0)g(0)^2}x_0\right)^2
      +\frac{p(0){\it\Delta}}{2{\it\Delta}+p(0)g(0)^2}x_0^2.
    \end{aligned}
  \end{equation}
\end{theorem}
\proof
  By condition \eqref{ftiao} and Proposition \ref{Prop3.1}, the mean-variance problem \eqref{question} is feasible for any $z\in \mathbb{R}$. By Theorem \ref{Thm5.1}, for every $\lambda\in \mathbb{R}$, the problem \eqref{unquestion} has a finite optimal value. Particularly, the problem \eqref{question} without the constraint $\mathbb{E}[x(\tau \wedge T)]=z$ is just the problem \eqref{unquestion} with $\lambda=0$, consequently it has a finite optimal value. Hence, \eqref{question} is finite for every $z\in \mathbb{R}$. Since $J_{MV}(\pi(\cdot))$ is strictly convex in $\pi(\cdot)$ and the constraint $J_1(\pi(\cdot))-z$ is affine in $\pi(\cdot)$, it follows from the duality theorem (see \cite{Luenberger1997} p.224 Theorem 1) that for any $z\in\mathbb{R}$, the optimal value of \eqref{question} is
  \begin{equation}\label{fitness}
    J^*_{MV}=\sup_{\lambda\in \mathbb{R}}\inf_{\pi(\cdot)\in \mathcal{U}}J(\pi(\cdot),\lambda)>-\infty.
  \end{equation}
  From Theorem \ref{Thm5.1}, $\inf_{\pi(\cdot)\in \mathcal{U}}J(\pi(\cdot),\lambda)$ is a quadratic function (see \eqref{optimalcost}) in $\lambda-z$. From the finiteness of the supremum value of this quadratic function, we have
  \begin{equation*}
    \frac{1}{2}p(0)g(0)^2+{\it\Delta}-1\leq 0.
  \end{equation*}
  But if $\frac{1}{2}p(0)g(0)^2+{\it\Delta}-1=0$, then due to \eqref{optimalcost} and \eqref{fitness}, we must have $p(0)g(0)-2z=0$, which is a contradiction for any $z\in \mathbb{R}$. So we prove \eqref{quadratic}.

  On the other hand, we maximize the function \eqref{optimalcost} over $\lambda-z$. The optimal Lagrange multiplier $\lambda^*$ is given by \eqref{optimallambda}, and the optimal value of the variance ${\rm Var } x^*(\tau\wedge T)$ is given by \eqref{optimalvalue}. Finally, the optimal control \eqref{opcontrol} is obtained by \eqref{optimalcontrol} with $\lambda=\lambda^*$.  \rulex

The expression \eqref{optimalvalue} reveals explicitly the tradeoff between the reward and the risk at the terminal time. Different from the case without Markov chain \cite{Lim2002} and the case with deterministic exit time \cite{Zhou2000}, the efficient frontier in the present case with Markov chain and random exit time is not a perfect square. As a consequence, the agent is not able to achieve a risk-free investment. Owing to the influence of the exit timing risk and the independence between Markov chain and Brownian motion, the interest rate risk cannot be perfectly hedged by any portfolio consisting of the bank account and stocks. Then we give the minimum variance as follows, which can be achieved by a feasible portfolio.

\begin{corollary}[minimum variance]\label{6.2}
    Let Assumptions \ref{2.7}-\ref{2.9} and \eqref{ftiao} hold. The minimum terminal variance is
    \begin{equation}\label{c}
      {\rm Var}x^*_{min}(\tau\wedge T)=\frac{p(0){\it\Delta}}{2{\it\Delta}+p(0)g(0)^2}x_0^2 \geq 0,
    \end{equation}
    which is achieved by the portfolio
    \begin{equation}\label{cc}
      \pi^*_{min}(t)=-\left(\theta(t)+\frac{{\it\Lambda}(t)}{p(t-)}\right)\sigma(t)^{-1}\left[x^*_{min}(t-)- z_{min}g(t-)\right]+z_{min}\sigma(t)^{-1}\eta(t).
    \end{equation}
    Moreover, the corresponding expected terminal wealth is given by
    \begin{equation}\label{ccc}
      z_{min}:=\frac{p(0)g(0)}{2{\it\Delta}+p(0)g(0)^2}x_0
    \end{equation}
    and the corresponding Lagrange multiplier $\lambda^*_{min}=0$.
\end{corollary}
\proof
  We can easily get \eqref{c} and \eqref{ccc} from the expression \eqref{optimalvalue}. From \eqref{optimallambda} and \eqref{ccc}, we calculate $\lambda^*_{min}=0$. Finally, \eqref{cc} follows from \eqref{opcontrol}. \rulex

\begin{remark}
  As same as Lim and Zhou \cite{Lim2002}, Zhou and Yin \cite{Zhou2003} and Yu \cite{Yu2013}, due to the minimum variance, the parameter $z$ can be restricted on the interval $z\in [z_{min},+\infty)$ when we study the efficient frontier to the mean-variance problem \eqref{question}.
\end{remark}

Now we give the so-called mutual fund theorem.
\begin{theorem}[mutual fund theorem]
  Let Assumptions \ref{2.7}-\ref{2.9} and \eqref{ftiao} hold. Let $\pi^*_{min}(\cdot)$ denote the minimum variance portfolio defined in Corollary \ref{6.2}, and $\pi^*_1(\cdot)$ is another given efficient portfolio corresponding to $z_1>z_{min}$. Then a portfolio $\pi^*(\cdot)$ is efficient if and only if there exists an $\beta\geq 0$ such that
  \begin{equation}\label{aa}
    \pi^*(t)=(1-\beta)\pi^*_{min}(t)+\beta\pi^*_1(t), \quad t\in [0,T].
  \end{equation}
\end{theorem}
\proof
  On one hand, we give the proof of the if part. For given $z_{min}$ and $z_1$, we write the expressions of the corresponding Lagrange multipliers and efficient portfolios:
  \begin{equation*}
    \begin{aligned}
      &\lambda^*_{min}=z_{min}+\frac{2z_{min}-p(0)g(0)x_0}{p(0)g(0)^2+2{\it\Delta}-2}, \quad \quad \lambda^*_{1}=z_{1}+\frac{2z_{1}-p(0)g(0)x_0}{p(0)g(0)^2+2{\it\Delta}-2},\\
      &\pi^*(t)=-\left(\theta(t)+\frac{{\it\Lambda}(t))}{p(t-)}\right)\sigma(t) ^{-1}\left[x^*_{min}(t-)+ (\lambda^*_{min}-z_{min})g(t-) \right]
      -(\lambda^*_{min}-z_{min})\sigma(t))^{-1}\eta(t)^{\rm T},\\
      &\pi^*(t)=-\left(\theta(t)+\frac{{\it\Lambda}(t)}{p(t-)}\right)\sigma(t) ^{-1}\left[x^*_{1}(t-)+ (\lambda^*_{1}-z_{1})g(t-) \right]
      -(\lambda^*_{1}-z_{1})\sigma(t)^{-1}\eta(t)^{\rm T},\\
    \end{aligned}
  \end{equation*}
  For any $\beta\geq 0$, we define $z=(1-\beta)z_{min}+\beta z_1$. By the linearities of $\lambda^*$ with respect to $z$, $x^*$ with respect to $\lambda^*$ and $\pi^*$ with respect to $x^*$, $\lambda^*$ and $z$, we have
  \begin{equation*}
    \begin{aligned}
      &\lambda^*=(1-\beta)\lambda^*_{min}+\beta \lambda^*_1,\\
      &x^*=(1-\beta)x^*_{min}+\beta x^*_1,\\
      &\pi^*=(1-\beta)\pi^*_{min}+\beta\pi^*_1.
    \end{aligned}
  \end{equation*}
  So a portfolio $\pi^*(\cdot)$ defined by \eqref{aa} is an efficient portfolio corresponding to $z=(1-\beta)z_{min}+\beta z_1$.

  On the other hand, we give the proof of the only if part. When $\pi^*(\cdot)$ is an efficient portfolio corresponding to a certain $z\geq z_{min}$, we can write $z=(1-\beta)z_{min}+\beta z_1$ with some $\beta \geq 0$. The above analysis leads to that $\pi^*(\cdot)$ must be in form of \eqref{aa}.   \rulex

\begin{remark}
  The mutual fund theorem implies that any efficient portfolio can be constructed as a combination of the minimum variance portfolio $\pi^*_{min}(\cdot)$ and another given efficient portfolio (called the mutual fund) $\pi^*_1(\cdot)$. In other words, the investor need only to invest in the minimum variance portfolio $\pi^*_{min}(\cdot)$ and the mutual fund $\pi^*_1(\cdot)$ to achieve the efficiency. The factor $\beta\geq 0$ means that the investor cannot short sell the mutual fund $\pi^*_1(\cdot)$.
\end{remark}

\begin{example}[numerical example]
Assume $r=0.1$, $\mu=0.3$, $\sigma=0.5$, $x_0=1$ and $T=1$. Then the efficient frontiers corresponding to the densities $f=0,\ 0.5,\ 0.8$ are plotted in Figure 1, respectively.
\end{example}

\begin{center}
  \centerline
  {\includegraphics[scale=0.6]{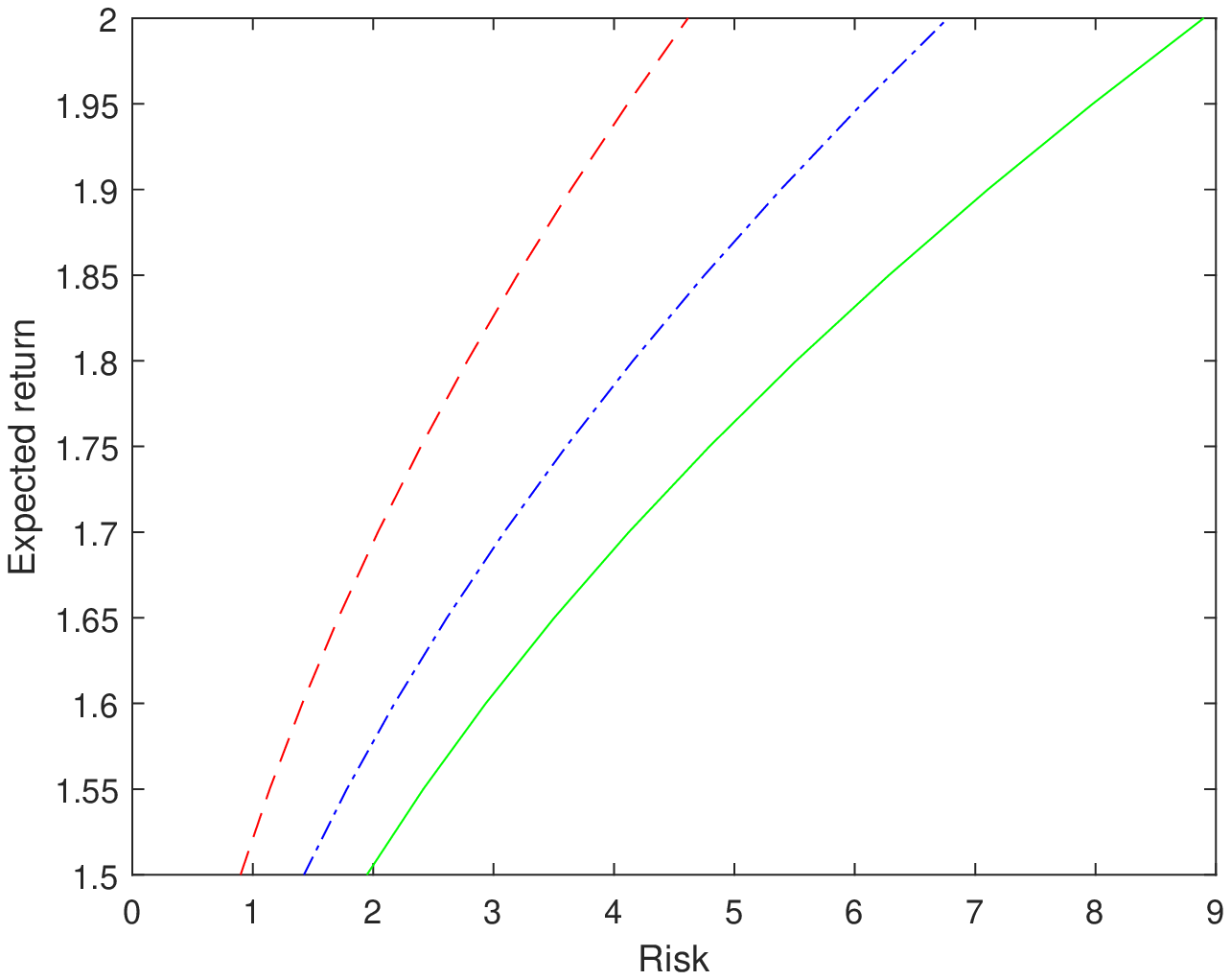}}\vskip3mm
\centering{\small {\bf Figure 1}\ \ Efficient frontier: f=0 (red), f=0.5 (blue) and f=0.8 (green) \label{fig1}}
\end{center}

Figure 1 shows that the larger the density of the random horizon is, the larger the variance of terminal wealth is. This is reasonable because a larger density implies that more extra uncertainty owing to the random horizon is added to the model, resulting in the increment of the overall risk.

\section{Conclusion}

In this paper, we have studied a continuous-time mean-variance portfolio selection problem under non-Markovian regime-switching model with random horizon. In our setting, an investor is uncertain about exit time. We assume that the exit time is $\tau \wedge T$, where $\tau$ is a $\mathcal{A}$-measurable positive random variable. It means that we have considered all uncertain factors affecting the investment, not just the price information. We use a submartingale to characterize the conditional distribution of random time $\tau$ and reconstruct the mean-variance problem according to the Doob-Mayer decomposition theorem and some assumptions. A key difficulty is to prove the global solvability of the stochastic Riccati equation. Using a truncation technique, we transform the corresponding SRE into a one-dimentional BSDE with quadratic growth. With the help of BMO martingales technique and comparison theorem, we proved the global solvability of the stochastic Riccati equation and the auxiliary regime-switching BSDE arising from the mean-variance problem.

For the mean-variance problem with regime-switching and random horizon, we obtained the efficient portfolios and efficient frontier in closed forms. Different from the case where the market is complete, the efficient frontier is no longer presented in a quadratic form. Due to the challenging structure of auxiliary regime-switching BSDE and ${\it\Delta}$, we obtained some results different from the existing works. We also proved that a risk-free investment is not able to be achieved. This phenomenon is reasonable because the random exit time and the Markov chain introduce new uncertainty, and such uncertainty can not be hedged by any portfolio consisting of the bond and stocks. In addition, we also present a minimum variance portfolio and a mutual fund theorem to illustrate.







\end{document}